\pdfoutput=1
\documentclass[12pt]{iopart}

\expandafter\let\csname equation*\endcsname\relax
\expandafter\let\csname endequation*\endcsname\relax
\usepackage[fleqn]{amsmath}
\usepackage{amssymb,amsthm,mathtools}
\usepackage{dsfont}
\usepackage{graphicx}
\usepackage{microtype}
\usepackage{cite}
\usepackage{xcolor}
\usepackage[colorlinks=true,allcolors={blue!60!black}]{hyperref}
\hypersetup{pdftitle={Majorization for superpositions of nonorthogonal states and the existence of golden states},pdfauthor={G\"{o}khan Torun}}
\newtheorem{theorem}{Theorem}
\newtheorem{proposition}{Proposition}
\newtheorem{lemma}{Lemma}
\newtheorem{corollary}{Corollary}

\newcommand{\ket}[1]{|#1\rangle}
\newcommand{\bra}[1]{\langle #1|}
\newcommand{\braket}[2]{\langle #1|#2\rangle}
\newcommand{\Id}{\mathds{1}}
\newcommand{\cperp}[1]{\ket{c_{#1}^{\perp}}}
\newcommand{\bS}{\mathbf{S}}
\newcommand{\bu}{\mathbf{u}}
\newcommand{\bsig}{\boldsymbol{\sigma}}
\newcommand{\bpsi}{\boldsymbol{\psi}}
\newcommand{\bphi}{\boldsymbol{\varphi}}
\newcommand{\bG}{\mathbf{G}}
\newcommand{\bP}{\mathbf{P}}
\newcommand{\bK}{\mathbf{K}}
\newcommand{\bD}{\mathbf{D}}
\newcommand{\bY}{\mathbf{Y}}
\newcommand{\bX}{\mathbf{X}}
\newcommand{\bff}{\mathbf{f}}
\newcommand{\MR}{\mathcal{M}_{\mathrm{R}}}
\newcommand{\MG}{\mathcal{M}_{\mathrm{G}}}
\newcommand{\Fc}{\mathcal{F}_{\mathcal{C}}}
\newcommand{\lmin}{\lambda_{\min}}
\newcommand{\T}{\mathsf{T}}

\begin{document}

\title{Majorization for superpositions of nonorthogonal states and the existence of golden states}

\author{G\"{o}khan Torun}

\address{Department of Physics, Faculty of Arts and Sciences, \.{I}zmir University of Economics, \.{I}zmir, 35330, Turkiye}

\ead{gtorung@gmail.com}

\begin{abstract}
Golden states, from which every state can be prepared by free operations, are the units of a resource theory. In the resource theory of superposition relative to linearly independent, nonorthogonal reference states, they were known only for two reference states and for special overlap matrices, all of them maximally coherent in the L\"{o}wdin basis. We determine what decides their existence and when this L\"{o}wdin pattern holds. For every overlap matrix, the robustness and the geometric measure of superposition are maximized by the same states, which overlap equally with all reference states; the maximal values are set by the ground-state energy of a classical $XY$ model on the dual basis. These states, and only these, are golden under the maximal superposition-free operations, but they need not be maximally coherent in the L\"{o}wdin basis. For the Kraus-type free operations, we prove Gram majorization, a necessary and sufficient criterion for converting pure states into targets with nonzero coefficients, which reduces to majorization for orthonormal reference states. It shows that every golden state of these operations is maximally coherent in the L\"{o}wdin basis, yet golden states exist only for special overlap matrices: three reference states that are not orthonormal admit one if and only if their overlaps have a common modulus $s$ and their Bargmann phase $\Phi$ obeys $s\geq2\cos[(\pi+|\Phi|)/3]$, so that a geometric phase decides whether the theory has a unit. Finally, equal-overlap states yield a complete family of superposition witnesses in the L\"{o}wdin basis, which show that white noise is a resource unless the reference states are orthonormal.
\end{abstract}



\section{Introduction}
\label{Sec:Intro}

Quantum resource theories treat features that rest on the superposition principle, such as coherence and entanglement, as resources~\cite{Chitambar2019QRTs}, and in a broad class of them, majorization decides which states can be converted into which~\cite{Torun2023Review}. The resource theory of coherence (RTC) concerns superposition with respect to a fixed orthonormal basis: its free states, the incoherent states, are the mixtures of the basis states~\cite{Baumgratz2014QCo,Winter2016,Streltsov2017QCo}. In many physical settings, however, the natural reference states are not orthogonal, as for the Glauber coherent states of a bosonic mode or the valence-bond structures of quantum chemistry~\cite{Gallup2002VB}. Superposition relative to coherent states characterizes nonclassicality~\cite{Tan2017}, and superpositions of phase-shifted coherent states form the cat states of bosonic codes~\cite{Mirrahimi2014Cat}. Theurer \textit{et al}~\cite{Theurer2017QSup} formulated the resource theory of superposition (RTS) for this situation. Its free pure states are \(d\) reference states that are normalized and linearly independent but may have arbitrary overlaps; for orthonormal reference states, the RTS reduces to the RTC.

A central object of a resource theory is its unit, a state from which every other state can be prepared by free operations. In the RTC, this is the maximally coherent state, which is unique up to the phases of its coefficients~\cite{Peng2016MaxCoh} and from which every state can be obtained by incoherent operations~\cite{Baumgratz2014QCo}. Its counterpart in the RTS is a golden state~\cite{Theurer2017QSup,Torun2021StateTrans,Senyasa2022Golden}, and whether golden states exist depends on the class of free operations. For the superposition-free operations (FO) of~\cite{Theurer2017QSup}, whose Kraus operators individually map superposition-free states to superposition-free states up to normalization, golden states always exist for two reference states~\cite{Theurer2017QSup,Torun2021StateTrans}. For \(d\geq3\), they do not exist in general, as the example of three states with equal overlap \(1/2\) shows~\cite{Theurer2017QSup}, and they were found only for special overlap matrices, such as equal negative overlaps~\cite{Senyasa2022Golden}. Using L\"{o}wdin symmetric orthogonalization (LSO)~\cite{Lowdin1950,CarlsonKeller1957,Aiken1980}, it was observed in~\cite{Torun2023Golden} that all known golden states are maximally coherent in the L\"{o}wdin basis, the orthonormal basis closest to the reference states in the least-squares sense. If this pattern held in general, golden states would be the maximally coherent states of a single orthonormal basis, as in the RTC.

This observation raises three questions. Does maximal coherence in the L\"{o}wdin basis characterize maximal superposition, or only the golden states of a particular class of free operations? Which overlap matrices admit golden states, and what decides their existence? How can superposition be certified in the L\"{o}wdin basis, given that the superposition-free states are themselves coherent in this basis~\cite{Torun2023Golden}? We answer all three. The L\"{o}wdin pattern turns out to be a property of the free operations, not of maximal superposition, and a geometric phase of the reference states can decide whether golden states exist.

First, let \(\bS\) be the overlap matrix of the reference states \(\ket{c_j}\), with entries \(S_{jk}=\braket{c_j}{c_k}\). For every \(\bS\), the robustness and the geometric measure of superposition are maximized by the same pure states (theorem~\ref{Thm:Maximal}): the superpositions \(\sum_ju_j\cperp{j}\) of the dual basis vectors, defined by \(\braket{c_j}{c_k^{\perp}}=\delta_{jk}\), with coefficients \(u_j\) of unit modulus and optimal phases. These states overlap equally with all reference states, and the maximal values of both measures follow from the single number
\begin{equation}\label{Eq:GammaIntro}
\Gamma(\bS)\coloneqq\max_{|u_j|=1}\bu^{\dagger}\bS^{-1}\bu ,
\end{equation}
which is also the largest advantage that any state offers over the superposition-free states in channel discrimination. This number is set by the ground-state energy of a classical \(XY\) model whose couplings are the overlaps of the dual basis vectors. Frustration in this model can force the maximizing states to have complex coefficients even when all overlaps are real, and \(\Gamma\) depends on the geometric phases of the reference states, not only on the moduli of their overlaps. Under the maximal superposition-free operations (MFO) of~\cite{Theurer2017QSup}, which comprise all channels that map superposition-free states to superposition-free states, these states, and only these, are golden for every \(\bS\). They are maximally coherent in the L\"{o}wdin basis if and only if \(\Gamma=d/\lmin\), with \(\lmin\) the smallest eigenvalue of \(\bS\) (proposition~\ref{Prop:Lowdin}).

Second, for FO, we prove Gram majorization (theorem~\ref{Thm:Gram}). A pure state \(\ket{\psi}=\sum_j\psi_j\ket{c_j}\) can be converted deterministically into a target \(\ket{\varphi}=\sum_j\varphi_j\ket{c_j}\) with nonzero coefficients \(\varphi_j\) if and only if the Gram matrices of their components \(\psi_j\ket{c_j}\) and \(\varphi_j\ket{c_j}\) obey a semidefinite ordering up to a mixture of permutations. For orthonormal reference states, this reduces to the majorization criterion of the RTC~\cite{Du2015,Du2017Erratum,Winter2016,ChitambarGour2016PRA,ChitambarGour2017Erratum,Zhu2017}, whereas earlier conversion conditions in the RTS assumed real coefficients and real overlaps~\cite{Torun2021StateTrans}. Gram majorization establishes for all of FO the necessary conditions for golden states that were derived in~\cite{Senyasa2022Golden} for a restricted family of Kraus operators, and it thereby proves that every golden state under FO is maximally coherent in the L\"{o}wdin basis. These conditions are not sufficient, and Gram majorization decides existence in new cases (theorem~\ref{Thm:GoldenFO} and corollary~\ref{Cor:Three}). If \(\bS\) has exactly two distinct eigenvalues, at least one of them nondegenerate, golden states under FO exist if and only if \(\lmin\) is nondegenerate; in particular, they never exist for \(d\geq3\) reference states with equal positive overlaps. Three reference states that are not orthonormal admit golden states under FO if and only if their overlaps have a common modulus \(s\) and their Bargmann phase \(\Phi=\arg(S_{12}S_{23}S_{31})\) obeys
\begin{equation}\label{Eq:CriterionIntro}
s\geq2\cos\frac{\pi+|\Phi|}{3} .
\end{equation}
Between the real cases \(\Phi=0\), which never admits golden states, and \(|\Phi|=\pi\), which always does, the geometric phase thus decides whether the theory has a unit.

Third, the vectors \(\sum_ju_j\cperp{j}\), now with arbitrary phases, yield a complete family of superposition witnesses (proposition~\ref{Prop:Witness}). When \(\Gamma=d/\lmin\), all superposition-free states carry the same L\"{o}wdin coherence weighted with the optimal phases, \(\lmin-1\). Any deviation from this value certifies superposition, and a positive excess bounds the robustness from below. Each witness requires a single two-outcome measurement, which, for phase-shifted coherent states, is a measurement of the photon number modulo \(d\). As a consequence, the maximally mixed state on the span of the reference states, which is free in the RTC, is a superposition resource, with robustness at least \(\Gamma/d-1\), whenever the reference states are not orthonormal.

Section~\ref{Sec:Setting} sets up the notation, the free operations, and the two superposition measures. Section~\ref{Sec:Maximal} characterizes maximal superposition and the golden states under MFO, section~\ref{Sec:Gram} proves Gram majorization, and section~\ref{Sec:GoldenFO} applies it to golden states under FO. Section~\ref{Sec:Witness} constructs the witnesses, and section~\ref{Sec:Conclusion} concludes.

\section{Setting}
\label{Sec:Setting}

This section fixes the notation, collects the properties of nonorthogonal bases that we use, and recalls the free states, free operations, and measures of the RTS. Throughout, operators on a Hilbert space carry a hat, and bold symbols denote vectors in \(\mathbb{C}^d\) and \(d\times d\) matrices; in particular, \(\hat{\Id}\) is the identity operator and \(\mathbf{1}\) is the identity matrix.

\subsection{Reference, L\"{o}wdin, and dual bases}
\label{Sec:Bases}

Let \(\mathcal{C}=\{\ket{c_j}\}_{j=1}^{d}\), with \(d\geq2\), be a set of normalized reference states that are linearly independent but not necessarily orthogonal, and let \(\mathcal{H}\) be the \(d\)-dimensional Hilbert space they span. Thus \(\mathcal{C}\) is a basis of \(\mathcal{H}\); in the RTS, its elements are the pure superposition-free states~\cite{Theurer2017QSup}. All states and operators considered below act on \(\mathcal{H}\). The overlap (Gram) matrix \(\bS\) of \(\mathcal{C}\), with entries \(S_{ij}\coloneqq\braket{c_i}{c_j}\), is Hermitian with unit diagonal, and it is positive definite because the reference states are linearly independent. Hence \(\bS\) has orthonormal eigenvectors \(\mathbf{v}_k\) with positive eigenvalues \(\lambda_k\), and
\begin{equation}\label{Eq:Spectral}
\bS=\sum_{k=1}^{d}\lambda_k\mathbf{v}_k\mathbf{v}_k^{\dagger}, \quad \bS^{\pm1/2}\coloneqq\sum_{k=1}^{d}\lambda_k^{\pm1/2}\mathbf{v}_k\mathbf{v}_k^{\dagger},
\end{equation}
where \(\bS^{1/2}\) is the unique positive definite square root of \(\bS\) and \(\bS^{-1/2}\) is its inverse. Since \(\bS\) has unit diagonal, its eigenvalues sum to \(\Tr\bS=d\), so the smallest one, \(\lmin\), obeys \(0<\lmin\leq1\), with \(\lmin=1\) if and only if the reference states are orthonormal.

The L\"{o}wdin basis \(\mathcal{L}=\{\ket{l_j}\}_{j=1}^{d}\) of \(\mathcal{C}\) is the orthonormal basis obtained by LSO~\cite{Lowdin1950},
\begin{equation}\label{Eq:Lowdin}
\ket{l_j}\coloneqq\sum_{k=1}^{d}\left[\bS^{-1/2}\right]_{kj}\ket{c_k}.
\end{equation}
In contrast to sequential orthogonalization schemes such as Gram--Schmidt, LSO singles out no reference state: relabeling the reference states relabels the L\"{o}wdin states accordingly, and every unitary that permutes the reference states permutes the L\"{o}wdin states in the same way. Moreover, \(\mathcal{L}\) is the orthonormal basis closest to \(\mathcal{C}\) in the least-squares sense: among all orthonormal bases \(\{\ket{o_j}\}_{j=1}^{d}\) of \(\mathcal{H}\), it alone minimizes the sum of squared distances \(\sum_j\|\,\ket{c_j}-\ket{o_j}\,\|^2\)~\cite{CarlsonKeller1957,Aiken1980}.

Besides \(\mathcal{L}\), we use the dual basis \(\mathcal{C}^{\perp}=\{\cperp{j}\}_{j=1}^{d}\) of \(\mathcal{H}\), which is uniquely determined by the biorthogonality condition \(\braket{c_i}{c_j^{\perp}}=\delta_{ij}\). Inverting equation~\eqref{Eq:Lowdin} and imposing this condition gives the components of the reference states and of the dual basis vectors in \(\mathcal{L}\),
\begin{equation}\label{Eq:Bases}
\braket{l_k}{c_j}=\left[\bS^{1/2}\right]_{kj}, \quad \braket{l_k}{c_j^{\perp}}=\left[\bS^{-1/2}\right]_{kj}.
\end{equation}
Combined with equation~\eqref{Eq:Lowdin}, these components give
\begin{equation}\label{Eq:Biorth}
\cperp{j}=\sum_{k=1}^{d}\left[\bS^{-1}\right]_{kj}\ket{c_k}, \quad \braket{c_i^{\perp}}{c_j^{\perp}}=\left[\bS^{-1}\right]_{ij},
\end{equation}
and the resolutions of the identity \(\hat{\Id}=\sum_j\ket{c_j}\!\bra{c_j^{\perp}}=\sum_j\cperp{j}\!\bra{c_j}\) on \(\mathcal{H}\). Since the reference states are normalized, the Cauchy--Schwarz inequality and equation~\eqref{Eq:Biorth} give \([\bS^{-1}]_{jj}=\braket{c_j^{\perp}}{c_j^{\perp}}\geq|\braket{c_j}{c_j^{\perp}}|^2=1\) for every \(j\).

The uniform mixture of the reference states,
\begin{equation}\label{Eq:Chi}
\hat{\chi}\coloneqq\frac{1}{d}\sum_{j=1}^{d}\ket{c_j}\!\bra{c_j},
\end{equation}
has the matrix \(\bS/d\) in \(\mathcal{L}\), by equation~\eqref{Eq:Bases}. Hence its eigenvalues are \(\lambda_k/d\), its eigenvectors, written in \(\mathcal{L}\), are the \(\mathbf{v}_k\), and its off-diagonal elements in \(\mathcal{L}\) are the overlaps divided by \(d\). In particular, \(\hat{\chi}\) is diagonal in \(\mathcal{L}\) if and only if the reference states are orthonormal, in which case \(\mathcal{L}\) coincides with \(\mathcal{C}\). We call an eigenstate of \(\hat{\chi}\) with the smallest eigenvalue, \(\lmin/d\), a ground state of \(\hat{\chi}\); by theorem~\ref{Thm:GoldenFO}(a), every golden state under FO (section~\ref{Sec:Free}) is such a ground state.

\subsection{States and weights}
\label{Sec:States}

By the first resolution of the identity, every state \(\hat{\sigma}\) on \(\mathcal{H}\) can be expanded in the reference states,
\begin{equation}\label{Eq:State}
\hat{\sigma}=\sum_{i,j=1}^{d}\sigma_{ij}\ket{c_i}\!\bra{c_j}, \quad \sigma_{ij}\coloneqq\bra{c_i^{\perp}}\hat{\sigma}\cperp{j} ,
\end{equation}
and we call the matrix \(\bsig\) with entries \(\sigma_{ij}\) the coefficient matrix of \(\hat{\sigma}\). By equation~\eqref{Eq:Bases}, the matrix of \(\hat{\sigma}\) in \(\mathcal{L}\) is the L\"{o}wdin-transformed coefficient matrix \(\bsig_{\mathrm{L}}\), whose diagonal elements are the L\"{o}wdin weights \(w_j\),
\begin{equation}\label{Eq:LowdinTransform}
\bsig_{\mathrm{L}}=\bS^{1/2}\bsig\bS^{1/2}, \quad w_j\coloneqq\left[\bsig_{\mathrm{L}}\right]_{jj} .
\end{equation}
We call the off-diagonal elements of \(\bsig_{\mathrm{L}}\), which are the coherences of \(\hat{\sigma}\) in \(\mathcal{L}\), the L\"{o}wdin coherences. Since \(\bS^{1/2}\) is invertible, equation~\eqref{Eq:LowdinTransform} shows that \(\hat{\sigma}\succeq0\) if and only if \(\bsig\succeq0\), and that \(\Tr\hat{\sigma}=\Tr\bsig_{\mathrm{L}}=\Tr(\bsig\bS)\). The states are therefore exactly the operators whose coefficient matrices obey \(\bsig\succeq0\) and \(\Tr(\bsig\bS)=1\).

The L\"{o}wdin weights have a direct operational meaning. The square-root, or `pretty good', measurement for discriminating the reference states, each prepared with probability \(1/d\)~\cite{HausladenWootters1994,Hausladen1996,EldarForney2001}, has the measurement operators
\begin{equation}\label{Eq:SRM}
\hat{E}_j\coloneqq\frac{1}{d}\,\hat{\chi}^{-1/2}\ket{c_j}\!\bra{c_j}\hat{\chi}^{-1/2} .
\end{equation}
Since \(\hat{\chi}\) has the matrix \(\bS/d\) in \(\mathcal{L}\), the operator \(\hat{\chi}^{-1/2}\) has the matrix \(\sqrt{d}\,\bS^{-1/2}\), and equation~\eqref{Eq:Bases} gives
\begin{equation}\label{Eq:SRMvector}
\hat{\chi}^{-1/2}\ket{c_j}=\sqrt{d}\sum_{k=1}^{d}\left[\bS^{-1/2}\bS^{1/2}\right]_{kj}\ket{l_k}=\sqrt{d}\,\ket{l_j} .
\end{equation}
Hence \(\hat{E}_j=\ket{l_j}\!\bra{l_j}\): the square-root measurement is the projective measurement in \(\mathcal{L}\), and applied to a state \(\hat{\sigma}\), it returns the guess \(\ket{c_j}\) with probability \(\bra{l_j}\hat{\sigma}\ket{l_j}=w_j\). This identification also follows from the least-squares property of \(\mathcal{L}\), because the square-root measurement is the least-squares measurement~\cite{EldarForney2001}. Furthermore, for equally likely reference states, the measurement in \(\mathcal{L}\) minimizes the error probability when the reference states are symmetric, as phase-shifted coherent states are~\cite{Ban1997}, and, more generally, when they form a single orbit of an abelian group of unitaries~\cite{EldarForney2001}.

For a pure state \(\ket{\psi}=\sum_j\psi_j\ket{c_j}\), with the coefficients \(\psi_j=\braket{c_j^{\perp}}{\psi}\) and the coefficient vector \(\bpsi\coloneqq(\psi_1,\dots,\psi_d)^{\T}\), the coefficient matrix factorizes as \(\bsig=\bpsi\bpsi^{\dagger}\), and the normalization \(\Tr(\bsig\bS)=1\) becomes \(\bpsi^{\dagger}\bS\bpsi=1\). The overlaps of \(\ket{\psi}\) with the reference states and its L\"{o}wdin coefficients are
\begin{equation}\label{Eq:Overlaps}
\braket{c_j}{\psi}=\left[\bS\bpsi\right]_j, \quad \braket{l_j}{\psi}=\left[\bS^{1/2}\bpsi\right]_j .
\end{equation}
As in the RTC~\cite{Baumgratz2014QCo,Peng2016MaxCoh}, we call a pure state maximally coherent in \(\mathcal{L}\) if all its L\"{o}wdin coefficients have modulus \(1/\sqrt{d}\), that is, if its L\"{o}wdin weights are uniform. We also use the Chirgwin--Coulson (CC) weights
\begin{equation}\label{Eq:CC}
\tilde{\psi}_j\coloneqq\psi_j^{*}\braket{c_j}{\psi}, \quad \sum_{j=1}^{d}\tilde{\psi}_j=\braket{\psi}{\psi}=1,
\end{equation}
which extend those of valence-bond theory~\cite{Chirgwin1950,Gallup2002VB} from real to complex coefficients and overlaps, and we collect them in the vector \(\tilde{\bpsi}\coloneqq(\tilde{\psi}_1,\dots,\tilde{\psi}_d)^{\T}\). Unlike the L\"{o}wdin weights, the CC weights are in general not probabilities: they need not be real, and their real parts can be negative. For orthonormal reference states, however, both the L\"{o}wdin weights and the CC weights reduce to \(|\psi_j|^2\). CC weights enter earlier conditions for state conversion~\cite{Torun2021StateTrans} and for golden states~\cite{Senyasa2022Golden}, and they reappear in section~\ref{Sec:Gram} as the row sums of the Gram matrices that govern conversion.

\subsection{Free states, free operations, and golden states}
\label{Sec:Free}

The free, or superposition-free, states are the mixtures of the reference states, \(\hat{\delta}=\sum_jp_j\ket{c_j}\!\bra{c_j}\) with probabilities \(p_j\), that is, the states with a diagonal coefficient matrix; we denote their set by \(\Fc\)~\cite{Theurer2017QSup}. We consider two classes of free operations, both introduced in~\cite{Theurer2017QSup}. The first class, FO, consists of the channels that admit a Kraus decomposition into free Kraus operators \(\hat{K}_n\), which map each free state to a multiple of a free state. Such a Kraus operator maps every \(\ket{c_j}\) to a multiple of a single reference state, because \(\hat{K}_n\ket{c_j}\!\bra{c_j}\hat{K}_n^{\dagger}\) is proportional to a free state of rank at most one and the reference states are linearly independent. It therefore has the form~\cite{Theurer2017QSup}
\begin{equation}\label{Eq:FOKraus}
\hat{K}_n=\sum_{j=1}^{d}\kappa_{n,j}\ket{c_{f_n(j)}}\!\bra{c_j^{\perp}},
\end{equation}
with complex \(\kappa_{n,j}\) and functions \(f_n\colon\{1,\dots,d\}\to\{1,\dots,d\}\). On coefficient vectors, \(\hat{K}_n\) acts as \(\bpsi\mapsto\bK_n\bpsi\), where the matrix \([\bK_n]_{ij}\coloneqq\kappa_{n,j}\delta_{i,f_n(j)}\) has at most one nonzero entry in each column. Because \(\bra{\psi}\hat{K}_n^{\dagger}\hat{K}_n\ket{\psi}=(\bK_n\bpsi)^{\dagger}\bS(\bK_n\bpsi)\), trace preservation, \(\sum_n\hat{K}_n^{\dagger}\hat{K}_n=\hat{\Id}\), reads
\begin{equation}\label{Eq:Completeness}
\sum_n\bK_n^{\dagger}\bS\bK_n=\bS .
\end{equation}
Since each Kraus operator is free, an operation in FO maps free states to free states even when its outcome \(n\) is recorded.

The second class, MFO, consists of all channels \(\Lambda\) with \(\Lambda(\Fc)\subseteq\Fc\). Unlike FO, it requires only the channel as a whole to preserve \(\Fc\), and it strictly contains FO~\cite{Theurer2017QSup}. For \(\bS=\mathbf{1}\), FO and MFO reduce to the incoherent and the maximally incoherent operations of the RTC~\cite{Baumgratz2014QCo,Streltsov2017QCo}.

If another set \(\mathcal{C}'=\{\ket{c_j'}\}_{j=1}^{d}\) of reference states has the same overlap matrix, \(\braket{c_i'}{c_j'}=S_{ij}\), then \(\ket{c_j'}=\hat{U}\ket{c_j}\) for all \(j\), with a unitary \(\hat{U}\) from \(\mathcal{H}\) onto the span of \(\mathcal{C}'\). This unitary maps the free states and both classes of free operations of \(\mathcal{C}\) onto those of \(\mathcal{C}'\), so the resource-theoretic properties studied in this paper depend on \(\mathcal{C}\) only through \(\bS\). Moreover, rephasing the reference states, \(\ket{c_j}\to e^{i\eta_j}\ket{c_j}\) with real \(\eta_j\), changes neither the free states nor the free operations, whereas it replaces each overlap \(S_{ij}\) by \(e^{i(\eta_j-\eta_i)}S_{ij}\). These properties therefore depend on \(\bS\) only up to such rephasings.

A state is \emph{golden} for a class of operations if every state can be obtained from it deterministically by an operation of that class~\cite{Senyasa2022Golden}. Because FO is contained in MFO, a golden state under FO is also golden under MFO. For FO, the golden states are the states of maximal superposition of Theurer \textit{et al}~\cite{Theurer2017QSup}. In the general framework of Liu, Bu, and Takagi~\cite{LiuBuTakagi2019}, however, such states are called root states, and the term golden state refers to pure states that maximize resource measures. To avoid confusion, we reserve the term `maximally superposed' for the maximizers of the superposition measures introduced next.

\subsection{Two measures}
\label{Sec:Measures}

We quantify superposition by two measures. First, the robustness of superposition~\cite{Theurer2017QSup},
\begin{equation}\label{Eq:RobDef}
\MR(\hat{\sigma})\coloneqq\min_{\hat{\tau}}\left\{t\geq0:\ \frac{\hat{\sigma}+t\hat{\tau}}{1+t}\in\Fc\right\},
\end{equation}
where \(\hat{\tau}\) runs over all states, is the smallest relative weight of noise that must be mixed into \(\hat{\sigma}\) to make it free. Second, we define the geometric measure of superposition as
\begin{equation}\label{Eq:GeoDef}
\MG(\hat{\sigma})\coloneqq1-\max_{\hat{\delta}\in\Fc}F(\hat{\sigma},\hat{\delta}),
\end{equation}
with the fidelity \(F(\hat{\rho},\hat{\sigma})\coloneqq(\Tr|\sqrt{\hat{\rho}}\sqrt{\hat{\sigma}}|)^2\). For \(\bS=\mathbf{1}\), the two measures reduce to the robustness of coherence~\cite{Napoli2016,Piani2016} and the geometric coherence~\cite{Streltsov2015}.

Both measures vanish exactly on \(\Fc\) and are nonincreasing under MFO, and hence under FO. For \(\MR\), whose monotonicity under FO was shown in~\cite{Theurer2017QSup}, a channel \(\Lambda\) in MFO maps every decomposition \(\hat{\sigma}+t\hat{\tau}=(1+t)\hat{\delta}\) with \(\hat{\delta}\in\Fc\) to the decomposition \(\Lambda(\hat{\sigma})+t\Lambda(\hat{\tau})=(1+t)\Lambda(\hat{\delta})\) of \(\Lambda(\hat{\sigma})\), with the same \(t\) and with \(\Lambda(\hat{\delta})\in\Fc\)~\cite{Regula2018}. For \(\MG\), the fidelity does not decrease under channels~\cite{NielsenChuang2010}, and \(\Lambda\) maps \(\Fc\) into itself. The following lemma extends the corresponding results for coherence~\cite{Napoli2016,Piani2016,Streltsov2015}: it gives a semidefinite form of \(\MR\), which specializes the convex formulation of the robustness for general resource theories~\cite{Regula2018}, and closed expressions of both measures for pure states.

\begin{lemma}\label{Lem:Measures}
For every state \(\hat{\sigma}\) with coefficient matrix \(\bsig\),
\begin{equation}\label{Eq:RobSDP}
1+\MR(\hat{\sigma})=\min_{\bD\succeq\bsig}\Tr\bD=\max_{\bY\succeq0,\ Y_{jj}=1}\Tr(\bsig\bY),
\end{equation}
where \(\bD\) runs over the real diagonal matrices and \(\bY\) over the complex correlation matrices, that is, the positive semidefinite matrices with unit diagonal. For a pure state \(\ket{\psi}\) with coefficient vector \(\bpsi\), with \(\MR(\psi)\) short for \(\MR(\ket{\psi}\!\bra{\psi})\) and likewise for \(\MG\),
\begin{equation}\label{Eq:PureMeasures}
\begin{split}
1+\MR(\psi)&=\Bigl(\sum_{j}|\psi_j|\Bigr)^2,\\
1-\MG(\psi)&=\max_j|\braket{c_j}{\psi}|^2 .
\end{split}
\end{equation}
If all \(\psi_j\neq0\), the maximum in equation~\eqref{Eq:RobSDP} for \(\bsig=\bpsi\bpsi^{\dagger}\) is attained only at \(\bY=\bu\bu^{\dagger}\) with \(u_j=\psi_j/|\psi_j|\).
\end{lemma}

\begin{proof}
Let \(\hat{\sigma}+t\hat{\tau}=(1+t)\hat{\delta}\) with a state \(\hat{\tau}\) and a free state \(\hat{\delta}=\sum_jp_j\ket{c_j}\!\bra{c_j}\), as in equation~\eqref{Eq:RobDef}. In coefficient form, this decomposition reads
\begin{equation}\label{Eq:CoeffDecomp}
\bsig+t\boldsymbol{\tau}=\bD, \quad \bD\coloneqq(1+t)\,\mathrm{diag}(p_1,\dots,p_d),
\end{equation}
where \(\boldsymbol{\tau}\succeq0\) is the coefficient matrix of \(\hat{\tau}\). Hence \(\bD\succeq\bsig\) and \(\Tr\bD=1+t\). Conversely, let \(\bD\) be real diagonal with \(\bD\succeq\bsig\). Since \(S_{jj}=1\), \(\bD-\bsig\succeq0\), and \(\bS\succ0\),
\begin{equation}\label{Eq:TrD}
\Tr\bD=\Tr(\bD\bS)\geq\Tr(\bsig\bS)=1,
\end{equation}
with equality if and only if \(\bD=\bsig\). If \(\Tr\bD>1\), set \(t=\Tr\bD-1\). Then \(\boldsymbol{\tau}=(\bD-\bsig)/t\succeq0\) has \(\Tr(\boldsymbol{\tau}\bS)=1\), and \(\bD/(1+t)\succeq0\) is diagonal with unit trace, so they are the coefficient matrices of a state \(\hat{\tau}\) and a free state \(\hat{\delta}\) that satisfy equation~\eqref{Eq:CoeffDecomp}. If \(\Tr\bD=1\), then \(\bD=\bsig\) is diagonal, so \(\hat{\sigma}\) is itself free, which corresponds to \(t=0\). Every decomposition thus yields a feasible \(\bD\) with \(\Tr\bD=1+t\), and every feasible \(\bD\) yields a decomposition with \(t=\Tr\bD-1\), which proves the first equality in equation~\eqref{Eq:RobSDP}. As for coherence~\cite{Piani2016}, the maximization is the dual program of the minimization. Both programs are strictly feasible (take \(\bD=c\mathbf{1}\) with large \(c\), and \(\bY=\mathbf{1}\)), so by semidefinite-programming duality their optimal values coincide and both are attained, which proves the second equality.

For \(\bsig=\bpsi\bpsi^{\dagger}\), every correlation matrix \(\bY\) has nonnegative \(2\times2\) principal minors, \(1-|Y_{jk}|^2\geq0\). Hence
\begin{equation}\label{Eq:PureBound}
\Tr(\bsig\bY)=\sum_{j,k}\psi_j^{*}Y_{jk}\psi_k\leq\sum_{j,k}|\psi_j||\psi_k|=\Bigl(\sum_{j}|\psi_j|\Bigr)^2 .
\end{equation}
Equality holds for \(\bY=\bu\bu^{\dagger}\), \(u_j=\psi_j/|\psi_j|\) (any unimodular \(u_j\) if \(\psi_j=0\)), which proves the first line of equation~\eqref{Eq:PureMeasures}. If all \(\psi_j\neq0\), then, since the sum in equation~\eqref{Eq:PureBound} is real and each of its terms has modulus at most \(|\psi_j||\psi_k|\), equality requires \(\psi_j^{*}Y_{jk}\psi_k=|\psi_j||\psi_k|\) for all \(j,k\), that is, \(Y_{jk}=u_ju_k^{*}\), so this \(\bY\) is the only maximizer. Finally, the fidelity of \(\ket{\psi}\) with a free state \(\hat{\delta}\) is \(\bra{\psi}\hat{\delta}\ket{\psi}=\sum_jp_j|\braket{c_j}{\psi}|^2\), which is maximal when all weight is on a \(j\) with the largest \(|\braket{c_j}{\psi}|\); this proves the second line of equation~\eqref{Eq:PureMeasures}.
\end{proof}

\section{Maximal superposition}
\label{Sec:Maximal}

\subsection{Equal overlap with every reference state}
\label{Sec:EqualOverlap}

Maximally superposed states turn out to overlap equally with all reference states, that is, the modulus \(|\braket{c_j}{\psi}|\) does not depend on \(j\). This is the constant-overlap condition, introduced for general resource theories in~\cite{LiuBuTakagi2019} and applied to golden states of the RTS in~\cite{Senyasa2022Golden}. By the second resolution of the identity, every \(\ket{\psi}\in\mathcal{H}\) obeys \(\ket{\psi}=\sum_j\braket{c_j}{\psi}\cperp{j}\), so the states with this property are, up to normalization, the superpositions \(\sum_ju_j\cperp{j}\) with \(\bu\) in the set of phase vectors \(\mathbb{T}^d\coloneqq\{\bu\in\mathbb{C}^d:|u_j|=1\ \text{for all}\ j\}\). The number \(\Gamma(\bS)\) of equation~\eqref{Eq:GammaIntro}, which we abbreviate as \(\Gamma\), can be written as
\begin{equation}\label{Eq:Gamma}
\Gamma(\bS)=\max_{\bu\in\mathbb{T}^d}\bu^{\dagger}\bS^{-1}\bu=\max_{\bu\in\mathbb{T}^d}\Bigl\|\sum_{j=1}^{d}u_j\cperp{j}\Bigr\|^2 ,
\end{equation}
where the second form uses that \(\bS^{-1}\) is the overlap matrix of \(\mathcal{C}^{\perp}\), by equation~\eqref{Eq:Biorth}. We call \(\bu\) optimal if it attains the maximum, and we associate with it the normalized state
\begin{equation}\label{Eq:PsiU}
\ket{\Psi_{\bu}}\coloneqq\frac{1}{\sqrt{\Gamma}}\sum_{j=1}^{d}u_j\cperp{j} .
\end{equation}
Since \(\cperp{j}=\sum_k[\bS^{-1}]_{kj}\ket{c_k}\) by equation~\eqref{Eq:Biorth}, its coefficient vector and its overlaps with the reference states are
\begin{equation}\label{Eq:PsiUprops}
\boldsymbol{\Psi}_{\bu}=\frac{\bS^{-1}\bu}{\sqrt{\Gamma}}, \quad \braket{c_j}{\Psi_{\bu}}=\frac{u_j}{\sqrt{\Gamma}} .
\end{equation}
Thus \(\ket{\Psi_{\bu}}\) overlaps equally, with fidelity \(1/\Gamma\), with every reference state. For a general \(\bu\in\mathbb{T}^d\), the normalized state proportional to \(\sum_ju_j\cperp{j}\) has the common fidelity \(1/(\bu^{\dagger}\bS^{-1}\bu)\geq1/\Gamma\), so the optimal phases make this fidelity as small as possible. For \(\bS=\mathbf{1}\), every \(\bu\in\mathbb{T}^d\) is optimal, \(\Gamma=d\), and the states \(\ket{\Psi_{\bu}}\) are the maximally coherent states.

The optimal phases have a simple structure. If \(\bu\) is optimal, then changing a single phase \(u_j\) cannot increase \(\bu^{\dagger}\bS^{-1}\bu=[\bS^{-1}]_{jj}+2\,\mathrm{Re}\,(u_j^{*}z_j)+\mathrm{const}\), where \(z_j\coloneqq\sum_{k\neq j}[\bS^{-1}]_{jk}u_k\) and the constant does not depend on \(u_j\). Hence \(u_j^{*}z_j=|z_j|\), that is, \(z_j=|z_j|u_j\), and
\begin{equation}\label{Eq:Stationary}
\bS^{-1}\bu=\mathbf{R}\bu, \quad \mathbf{R}\coloneqq\mathrm{diag}(r_1,\dots,r_d),
\end{equation}
with \(r_j\coloneqq[\bS^{-1}]_{jj}+|z_j|\geq[\bS^{-1}]_{jj}\geq1\) and \(\sum_jr_j=\bu^{\dagger}\mathbf{R}\bu=\Gamma\). The first-order optimality condition of this maximization, familiar from phase synchronization~\cite{Boumal2016}, only requires \(\bS^{-1}\bu=\mathbf{R}\bu\) with some real diagonal \(\mathbf{R}\); at a maximum, in addition, every phase \(u_j\) is aligned with \(z_j\), so that \(r_j\geq[\bS^{-1}]_{jj}\). By equation~\eqref{Eq:PsiUprops}, \(\ket{\Psi_{\bu}}\) therefore has the coefficients \(r_ju_j/\sqrt{\Gamma}\), which are all nonzero, and the CC weights \(\tilde{\Psi}_j=r_j/\Gamma\), which are real and positive but in general not uniform.

\begin{theorem}[Maximal superposition]\label{Thm:Maximal}
For every set \(\mathcal{C}\) of reference states, the following hold.\\
(a) Every pure state obeys
\begin{equation}\label{Eq:Product}
\left[1+\MR(\psi)\right]\left[1-\MG(\psi)\right]\geq1,
\end{equation}
with equality if and only if all CC weights \(\tilde{\psi}_j\) are real and nonnegative and \(|\braket{c_j}{\psi}|=\max_k|\braket{c_k}{\psi}|\) for every \(j\) with \(\psi_j\neq0\).\\
(b) The maximum of \(\MR\) over all states is \(\Gamma-1\), and it is attained exactly by the states \(\ket{\Psi_{\bu}}\) with optimal \(\bu\). The maximum of \(\MG\) over all states is \(1-1/\Gamma\), and among pure states it is attained exactly by the same states.\\
(c) Every \(\ket{\Psi_{\bu}}\) with optimal \(\bu\) is golden under MFO, and every golden state under MFO is of this form.
\end{theorem}

\begin{proof}
(a) By equation~\eqref{Eq:CC},
\begin{equation}\label{Eq:ProductProof}
1=\sum_j\psi_j^{*}\braket{c_j}{\psi}\leq\sum_j|\psi_j||\braket{c_j}{\psi}|\leq\Bigl(\sum_j|\psi_j|\Bigr)\max_j|\braket{c_j}{\psi}| .
\end{equation}
Squaring and using equation~\eqref{Eq:PureMeasures} gives equation~\eqref{Eq:Product}, with equality if and only if both inequalities in equation~\eqref{Eq:ProductProof} are tight. The first is tight if and only if all \(\tilde{\psi}_j\) are real and nonnegative, because their sum is the positive number \(1\), and the second if and only if \(|\braket{c_j}{\psi}|\) is maximal for every \(j\) with \(\psi_j\neq0\).

(b) For a pure state, \(\sum_j|\psi_j|=\max_{\bu\in\mathbb{T}^d}|\bu^{\dagger}\bpsi|\), so equation~\eqref{Eq:PureMeasures} gives \(1+\MR(\psi)=\max_{\bu\in\mathbb{T}^d}|\bu^{\dagger}\bpsi|^2\). The Cauchy--Schwarz inequality gives
\begin{equation}\label{Eq:CS}
|\bu^{\dagger}\bpsi|^2=\bigl|\bigl(\bS^{-1/2}\bu\bigr)^{\dagger}\bigl(\bS^{1/2}\bpsi\bigr)\bigr|^2\leq\bigl(\bu^{\dagger}\bS^{-1}\bu\bigr)\bigl(\bpsi^{\dagger}\bS\bpsi\bigr)\leq\Gamma .
\end{equation}
Hence \(1+\MR(\psi)\leq\Gamma\) for every pure state. If \(1+\MR(\psi)=\Gamma\), choose \(u_j=\psi_j/|\psi_j|\) (any unimodular \(u_j\) if \(\psi_j=0\)). Then both inequalities in equation~\eqref{Eq:CS} are tight, so \(\bu\) is optimal and \(\bS^{1/2}\bpsi\propto\bS^{-1/2}\bu\), that is, \(\ket{\psi}=\ket{\Psi_{\bu}}\) up to a phase. Conversely, equation~\eqref{Eq:PsiUprops} gives \(1+\MR(\Psi_{\bu})\geq|\bu^{\dagger}\boldsymbol{\Psi}_{\bu}|^2=\Gamma\), so \(\MR(\Psi_{\bu})=\Gamma-1\).

By equation~\eqref{Eq:RobSDP}, \(\MR\) is convex, so its maximum over the convex set of states is attained at an extreme point, that is, at a pure state, and equals \(\Gamma-1\). Let \(\hat{\sigma}\) be any state with \(\MR(\hat{\sigma})=\Gamma-1\), and write \(\hat{\sigma}=\sum_kq_k\ket{\psi_k}\!\bra{\psi_k}\) with \(q_k>0\) and pure states \(\ket{\psi_k}\). Convexity gives \(\Gamma-1=\MR(\hat{\sigma})\leq\sum_kq_k\MR(\psi_k)\leq\Gamma-1\), so every \(\ket{\psi_k}\) has \(\MR(\psi_k)=\Gamma-1\) and equals some \(\ket{\Psi_{\bu_k}}\) up to a phase. Let \(\bY\) attain the maximum in equation~\eqref{Eq:RobSDP} for \(\hat{\sigma}\). Since \(\Gamma=\sum_kq_k\Tr(\boldsymbol{\Psi}_{\bu_k}\boldsymbol{\Psi}_{\bu_k}^{\dagger}\bY)\) and every term is at most \(\Gamma\), the same \(\bY\) attains the maximum for every \(\ket{\psi_k}\). The coefficients of \(\ket{\Psi_{\bu_k}}\) are all nonzero by equation~\eqref{Eq:Stationary}, so lemma~\ref{Lem:Measures} gives \(\bY=\bu_k\bu_k^{\dagger}\) for every \(k\). Hence all \(\bu_k\) coincide up to a global phase, and \(\hat{\sigma}=\ket{\Psi_{\bu_1}}\!\bra{\Psi_{\bu_1}}\).

For the geometric measure, part (a) gives \(1-\MG(\psi)\geq1/[1+\MR(\psi)]\geq1/\Gamma\) for every pure state, while \(1-\MG(\Psi_{\bu})=1/\Gamma\) by equation~\eqref{Eq:PsiUprops}. A pure state with \(\MG(\psi)=1-1/\Gamma\) must therefore have \(\MR(\psi)=\Gamma-1\), and hence be some \(\ket{\Psi_{\bu}}\). The maximum of \(\MG\) over all states is also attained at a pure state, because \(\sqrt{1-\MG(\hat{\sigma})}=\max_{\hat{\delta}\in\Fc}\Tr|\sqrt{\hat{\sigma}}\sqrt{\hat{\delta}}|\) is concave in \(\hat{\sigma}\), as the root fidelity is jointly concave~\cite{NielsenChuang2010} and \(\Fc\) is convex, and a concave function attains its minimum over the convex set of states at an extreme point.

(c) Let \(\hat{\sigma}'\) be any state. By equation~\eqref{Eq:RobDef} and part (b), there is a free state \(\hat{\delta}'\) with \(\hat{\sigma}'\preceq[1+\MR(\hat{\sigma}')]\hat{\delta}'\preceq\Gamma\hat{\delta}'\). As for coherence dilution under maximally incoherent operations~\cite{Zhao2018}, define the measure-and-prepare channel
\begin{equation}\label{Eq:MFOchannel}
\Lambda(\hat{X})\coloneqq\bra{\Psi_{\bu}}\hat{X}\ket{\Psi_{\bu}}\hat{\sigma}'+\Tr\bigl[\bigl(\hat{\Id}-\ket{\Psi_{\bu}}\!\bra{\Psi_{\bu}}\bigr)\hat{X}\bigr]\hat{\tau},
\end{equation}
where \(\hat{\tau}=(\Gamma\hat{\delta}'-\hat{\sigma}')/(\Gamma-1)\), with \(\Gamma\geq d\geq2\) by equation~\eqref{Eq:Bounds} below, is positive semidefinite with unit trace. By construction, \(\Lambda(\ket{\Psi_{\bu}}\!\bra{\Psi_{\bu}})=\hat{\sigma}'\). For every free state \(\hat{\delta}=\sum_jp_j\ket{c_j}\!\bra{c_j}\), equation~\eqref{Eq:PsiUprops} gives \(\bra{\Psi_{\bu}}\hat{\delta}\ket{\Psi_{\bu}}=1/\Gamma\), hence \(\Lambda(\hat{\delta})=\hat{\sigma}'/\Gamma+(1-1/\Gamma)\hat{\tau}=\hat{\delta}'\in\Fc\), and \(\Lambda\) is in MFO. Conversely, a golden state \(\hat{\sigma}_{\mathrm{g}}\) can be converted into \(\ket{\Psi_{\bu}}\), so monotonicity gives \(\MR(\hat{\sigma}_{\mathrm{g}})\geq\MR(\Psi_{\bu})=\Gamma-1\). By part (b), \(\hat{\sigma}_{\mathrm{g}}\) is therefore one of the states \(\ket{\Psi_{\bu}}\) with optimal \(\bu\).
\end{proof}

Theorem~\ref{Thm:Maximal} ties maximal superposition to equal overlap: a maximally superposed state lies as far as possible from all reference states at once, its largest fidelity with a reference state taking the smallest possible value \(1/\Gamma\). Part (a) is a trade-off between the two measures: a pure state with small robustness cannot be far from all reference states. The inequality itself holds in every convex resource theory, because \(\ket{\psi}\!\bra{\psi}\preceq[1+\MR(\psi)]\hat{\delta}\) for some free state \(\hat{\delta}\) implies \(1\leq[1+\MR(\psi)]\bra{\psi}\hat{\delta}\ket{\psi}\leq[1+\MR(\psi)][1-\MG(\psi)]\) (cf.~\cite{LiuBuTakagi2019}); its equality conditions are specific to the RTS.

In part (c), a golden state reaches every target through the measure-and-prepare channel~\eqref{Eq:MFOchannel}, which maps all free states to one and the same free state. This channel is in general not in FO; otherwise, every \(\ket{\Psi_{\bu}}\) would be golden under FO, which theorem~\ref{Thm:GoldenFO}(c) excludes for equal positive overlaps. Sections~\ref{Sec:Gram} and~\ref{Sec:GoldenFO} show what this restriction costs. Parts (b) and (c) also show that, under MFO, the two notions of Liu, Bu, and Takagi~\cite{LiuBuTakagi2019} coincide in the RTS: the states that maximize \(\MR\), which are also the pure states that maximize \(\MG\), are exactly the states from which every state can be prepared. The existence part of (c) is an instance of their root-state criterion, which applies because \(\Fc\) is the convex hull of the pure free states and \(\ket{\Psi_{\bu}}\) has the same overlap with every free state; specific to the RTS are the explicit form of these states for every \(\bS\) and the converse, that no other state is golden under MFO.

The number \(\Gamma\) also has a direct operational meaning. Because \(\Fc\) is convex and closed, the robustness of a state equals the largest advantage that it offers over all free states in channel discrimination~\cite{Takagi2019,TakagiRegula2019,Uola2019},
\begin{equation}\label{Eq:Operational}
1+\MR(\hat{\sigma})=\max_{\{\Lambda_i\},\{\hat{M}_i\}}\frac{p_{\mathrm{succ}}(\hat{\sigma})}{\max_{\hat{\delta}\in\Fc}p_{\mathrm{succ}}(\hat{\delta})} .
\end{equation}
Here the maximum runs over all instruments \(\{\Lambda_i\}\) and measurements \(\{\hat{M}_i\}\), and \(p_{\mathrm{succ}}(\hat{\rho})=\sum_i\Tr[\hat{M}_i\Lambda_i(\hat{\rho})]\) is the probability of identifying correctly which branch \(\Lambda_i\) of the instrument has acted. By theorem~\ref{Thm:Maximal}(b), \(\Gamma\) is therefore the largest discrimination advantage that any state can offer over the superposition-free states, and the states \(\ket{\Psi_{\bu}}\) with optimal \(\bu\) are the only states that achieve it. For \(\bS=\mathbf{1}\), this is the advantage \(d\) of maximally coherent states in phase discrimination~\cite{Napoli2016,Piani2016}.

\subsection{The L\"{o}wdin criterion}
\label{Sec:Lowdin}

Two elementary estimates bound \(\Gamma\). Averaging \(\bu^{\dagger}\bS^{-1}\bu\) over independent, uniformly distributed phases gives \(\Tr\bS^{-1}\), because the cross terms average to zero, and this average cannot exceed the maximum. The Rayleigh bound gives \(\bu^{\dagger}\bS^{-1}\bu\leq\|\bu\|^2/\lmin=d/\lmin\). Together with \(\Tr\bS^{-1}=\sum_k\lambda_k^{-1}\geq d^2/\sum_k\lambda_k=d\), which follows from the Cauchy--Schwarz inequality, these estimates give
\begin{equation}\label{Eq:Bounds}
d\leq\Tr\bS^{-1}\leq\Gamma\leq\frac{d}{\lmin} .
\end{equation}
Hence \(\Gamma=d\) if and only if \(\bS=\mathbf{1}\), because \(\Tr\bS^{-1}=d\) requires equal eigenvalues, which then all equal one. Nonorthogonal reference states therefore always allow a larger maximal robustness than the value \(d-1\) of the RTC. The upper bound decides whether the maximally superposed states are the maximally coherent ground states of \(\hat{\chi}\) found as golden states in~\cite{Senyasa2022Golden,Torun2023Golden}.

\begin{proposition}[L\"{o}wdin criterion]\label{Prop:Lowdin}
(a) \(\Gamma=d/\lmin\) if and only if \(\hat{\chi}\) has a ground state that is maximally coherent in \(\mathcal{L}\). In this case, the states \(\ket{\Psi_{\bu}}\) with optimal \(\bu\) are exactly the ground states of \(\hat{\chi}\) that are maximally coherent in \(\mathcal{L}\), with L\"{o}wdin coefficients \(\braket{l_j}{\Psi_{\bu}}=u_j/\sqrt{d}\).\\
(b) For every optimal \(\bu\), the following are equivalent: the L\"{o}wdin weights of \(\ket{\Psi_{\bu}}\) are uniform; its CC weights are uniform; \(\ket{\Psi_{\bu}}\) is an eigenstate of \(\hat{\chi}\).
\end{proposition}

\begin{proof}
(a) By the Rayleigh bound, \(\bu\in\mathbb{T}^d\) attains the upper bound in equation~\eqref{Eq:Bounds} if and only if \(\bS\bu=\lmin\bu\). Since \(\hat{\chi}\) has the matrix \(\bS/d\) in \(\mathcal{L}\), such a \(\bu\) exists if and only if \(\hat{\chi}\) has a ground state that is maximally coherent in \(\mathcal{L}\), namely the state with L\"{o}wdin coefficients \(u_j/\sqrt{d}\); this proves the first claim. If \(\Gamma=d/\lmin\), every optimal \(\bu\) obeys \(\bS\bu=\lmin\bu\), and equations~\eqref{Eq:Overlaps} and~\eqref{Eq:PsiUprops} give the L\"{o}wdin coefficients \(\bS^{-1/2}\bu/\sqrt{\Gamma}=\bu/\sqrt{d}\) of \(\ket{\Psi_{\bu}}\). Conversely, a ground state of \(\hat{\chi}\) with L\"{o}wdin coefficients \(u_j/\sqrt{d}\), \(\bu\in\mathbb{T}^d\), has \(\bS\bu=\lmin\bu\) and hence \(\bu^{\dagger}\bS^{-1}\bu=d/\lmin=\Gamma\). Thus \(\bu\) is optimal, and the state is \(\ket{\Psi_{\bu}}\).

(b) By equations~\eqref{Eq:Overlaps} and~\eqref{Eq:PsiUprops}, the L\"{o}wdin coefficients of \(\ket{\Psi_{\bu}}\) are \(\bS^{-1/2}\bu/\sqrt{\Gamma}\), whose squared moduli sum to one. If they have uniform moduli, then \(\bu'\coloneqq\sqrt{d/\Gamma}\,\bS^{-1/2}\bu\) lies in \(\mathbb{T}^d\), and by equation~\eqref{Eq:Stationary}
\begin{equation}\label{Eq:PropLowdinProof}
\bu'^{\dagger}\bS^{-1}\bu'=\frac{d}{\Gamma}\|\bS^{-1}\bu\|^2=\frac{d}{\Gamma}\sum_jr_j^2\geq\frac{1}{\Gamma}\Bigl(\sum_jr_j\Bigr)^2=\Gamma,
\end{equation}
with equality only if all \(r_j\) are equal. Since \(\bu'\in\mathbb{T}^d\), its value cannot exceed \(\Gamma\), so equality holds and \(\bS^{-1}\bu=(\Gamma/d)\bu\). The L\"{o}wdin coefficients, proportional to \(\bu\), then form an eigenvector of \(\bS\), so \(\ket{\Psi_{\bu}}\) is an eigenstate of \(\hat{\chi}\). Conversely, if \(\ket{\Psi_{\bu}}\) is an eigenstate of \(\hat{\chi}\), then its L\"{o}wdin coefficients \(\bS^{-1/2}\bu/\sqrt{\Gamma}\) form an eigenvector of \(\bS\), so \(\bS\bu\propto\bu\), all \(r_j\) are equal, and both the L\"{o}wdin coefficients \(\bS^{-1/2}\bu/\sqrt{\Gamma}\) and the CC weights \(r_j/\Gamma\) are uniform. Finally, uniform CC weights mean equal \(r_j\), that is, \(\bS^{-1}\bu\propto\bu\), which is again the eigenvector condition.
\end{proof}

Proposition~\ref{Prop:Lowdin} places the L\"{o}wdin picture of golden states~\cite{Torun2023Golden} in context. Maximal superposition and maximal L\"{o}wdin coherence are different properties. The maximally superposed states always overlap equally with the reference states, but they are maximally coherent ground states of \(\hat{\chi}\) only when \(\Gamma=d/\lmin\), that is, when the ground space of \(\hat{\chi}\) contains a state that is maximally coherent in \(\mathcal{L}\). The L\"{o}wdin weights of a maximally superposed state detect its eigenstate property: they are uniform exactly when the CC weights are, which ties together the two weight schemes of valence-bond theory~\cite{Gallup2002VB}.

Three examples illustrate the criterion. First, let \(d=2\) and \(S_{12}=se^{i\vartheta}\) with \(0<s<1\). The eigenvector of \(\bS\) with the smallest eigenvalue, \(\lmin=1-s\), is proportional to \((1,-e^{-i\vartheta})^{\T}\) and has uniform moduli. Hence
\begin{equation}\label{Eq:Qubit}
\Gamma=\frac{2}{1-s}, \quad \ket{\Psi_{\bu}}=\frac{\ket{c_1}-e^{-i\vartheta}\ket{c_2}}{\sqrt{2(1-s)}},
\end{equation}
which is the known golden state~\cite{Theurer2017QSup,Torun2021StateTrans,Senyasa2022Golden}.

Second, consider the phase-shifted coherent states \(\ket{\alpha\omega^{j}}\), \(j=1,\dots,d\), with \(\alpha\neq0\) and \(\omega=e^{2\pi i/d}\), used in cat codes~\cite{Mirrahimi2014Cat,Grimsmo2020}. Their overlaps \(S_{jk}=\exp[|\alpha|^2(\omega^{k-j}-1)]\) depend only on \(k-j\), so \(\bS\) is circulant, and the Fourier vectors \((\omega^{-mj})_{j=1}^{d}\), \(m=0,\dots,d-1\), are eigenvectors of \(\bS\) with uniform moduli. The eigenvalue of the \(m\)th Fourier vector is
\begin{equation}\label{Eq:CatEigen}
\lambda_m=d\,e^{-|\alpha|^2}\sum_{n\equiv m\ (\mathrm{mod}\ d)}\frac{|\alpha|^{2n}}{n!},
\end{equation}
and \(\lambda_m/d\) is the probability that the photon number of a coherent state with amplitude \(\alpha\) has residue \(m\) modulo \(d\). Hence \(\Gamma=d/\lmin\), and for every residue \(m\) that minimizes \(\lambda_m\), the cat state
\begin{equation}\label{Eq:Cat}
\ket{\Psi_{\bu}}\propto\sum_{j=1}^{d}\omega^{-mj}\ket{\alpha\omega^{j}}
\end{equation}
is maximally superposed. It contains only photon numbers \(n\equiv m\pmod d\), as is familiar from these codes, so the maximally superposed state lives on the least probable residue.

\begin{figure}[t]
\centering
\includegraphics[width=0.62\textwidth]{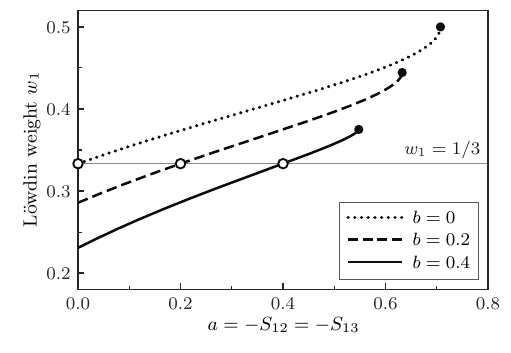}
\caption{L\"{o}wdin weight \(w_1\) of the maximally superposed state \(\ket{\Psi_{\bu}}\propto\sum_j\cperp{j}\) for three reference states with \(S_{12}=S_{13}=-a\) and \(S_{23}=-b\) as in equation~\eqref{Eq:Isosceles}, for \(b=0.4\) (solid), \(0.2\) (dashed), and \(0\) (dotted); the other two weights are \(w_2=w_3=(1-w_1)/2\). Each curve meets the line \(w_1=1/3\), on which \(\ket{\Psi_{\bu}}\) is maximally coherent in \(\mathcal{L}\), only at \(a=b\) (open circles), as proposition~\ref{Prop:Lowdin} requires. Each curve ends at \(a=\sqrt{(1-b)/2}\) (filled dots), where \(\lmin\to0\) and \(w_1\to(1-b)/(2-b)\).}
\label{Fig:LowdinWeight}
\end{figure}

Third, consider real nonpositive overlaps \(S_{12}=S_{13}=-a\) and \(S_{23}=-b\), with \(a,b\geq0\) and \(1-b-2a^2>0\), which makes \(\bS\) positive definite. Then \(\bS^{-1}\) is entrywise nonnegative (section~\ref{Sec:Frustration}), so equal phases are optimal, and \(\ket{\Psi_{\bu}}\propto\sum_j\cperp{j}\), with
\begin{equation}\label{Eq:Isosceles}
\Gamma=\frac{3+4a-b}{1-b-2a^2}, \quad \lmin=\frac{2-b-\sqrt{b^2+8a^2}}{2} .
\end{equation}
The vector \(\bu=(1,1,1)^{\T}\) is an eigenvector of \(\bS\) only for \(a=b\). Hence the criterion \(\Gamma=d/\lmin\) holds only there, and by proposition~\ref{Prop:Lowdin}(b) the L\"{o}wdin weights of \(\ket{\Psi_{\bu}}\) are uniform only there, as figure~\ref{Fig:LowdinWeight} shows.

\subsection{Frustration and geometric phases}
\label{Sec:Frustration}

Writing \(u_j=e^{i\theta_j}\) in equation~\eqref{Eq:Gamma} gives
\begin{equation}\label{Eq:XY}
\Gamma=\Tr\bS^{-1}+\max_{\theta_1,\dots,\theta_d}\,2\sum_{j<k}\bigl|\bigl[\bS^{-1}\bigr]_{jk}\bigr|\cos\bigl(\theta_k-\theta_j+\arg\bigl[\bS^{-1}\bigr]_{jk}\bigr).
\end{equation}
The maximized sum is minus the energy of a classical \(XY\) model~\cite{Villain1977,TeitelJayaprakash1983} with spins \(\theta_j\), whose couplings \(2|[\bS^{-1}]_{jk}|\) and bond phases \(\arg[\bS^{-1}]_{jk}\) are set by the dual overlaps \(\braket{c_j^{\perp}}{c_k^{\perp}}=[\bS^{-1}]_{jk}\); hence \(\Gamma-\Tr\bS^{-1}\) is minus its ground-state energy. A rephasing \(\ket{c_j}\to e^{i\eta_j}\ket{c_j}\), which leaves the resource theory unchanged (section~\ref{Sec:Free}), replaces \([\bS^{-1}]_{jk}\) by \(e^{i(\eta_k-\eta_j)}[\bS^{-1}]_{jk}\) and is absorbed by the shift \(\theta_j\to\theta_j-\eta_j\) of all angles. Hence \(\Gamma\) depends on the overlaps only through their moduli and their cyclic products \(S_{j_1j_2}S_{j_2j_3}\cdots S_{j_mj_1}\), the Bargmann invariants~\cite{Bargmann1964,SamuelBhandari1988,MukundaSimon1993}, whose phases are geometric phases. Bargmann invariants do not refer to any basis. They can be measured with generalized SWAP tests, and they witness coherence and imaginarity of sets of states~\cite{GalvaoBrod2020,Oszmaniec2024,Fernandes2024}.

\begin{figure}[t]
\centering
\includegraphics[width=0.62\textwidth]{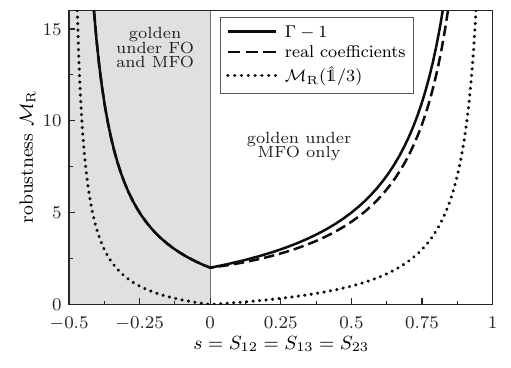}
\caption{Robustness of superposition for three reference states with equal real overlap \(s\). Solid: the maximum \(\Gamma-1\) over all states, attained by \(\sum_j\ket{c_j}\) (normalized) for \(s\leq0\) and by the chiral states~\eqref{Eq:Chiral} for \(s>0\). Dashed: the maximum over superpositions with real coefficients, equation~\eqref{Eq:RealTrine}, which falls below \(\Gamma-1\) for \(s>0\). Dotted: robustness of the maximally mixed state, \(\MR(\hat{\Id}/3)=\Gamma/3-1\), for which the bounds in equation~\eqref{Eq:WhiteNoise} coincide. All curves diverge at \(s=-1/2\) and \(s=1\), where \(\bS\) ceases to be positive definite. Golden states under FO exist exactly for \(s\leq0\) (theorem~\ref{Thm:GoldenFO}); golden states under MFO exist for all \(s\).}
\label{Fig:Trine}
\end{figure}

The \(XY\) model is unfrustrated if the angles can be chosen so that every bond is satisfied, that is, all rephased dual overlaps \(e^{-i\theta_j}[\bS^{-1}]_{jk}e^{i\theta_k}\) are nonnegative. Since the angles can absorb exactly those bond phases that add up to zero around every closed loop, this happens if and only if every cyclic product of nonzero dual overlaps, \([\bS^{-1}]_{j_1j_2}[\bS^{-1}]_{j_2j_3}\cdots[\bS^{-1}]_{j_mj_1}\), is positive. In general, the triangle inequality gives \(\Gamma\leq\sum_{j,k}|[\bS^{-1}]_{jk}|\), with equality exactly in the unfrustrated case, where
\begin{equation}\label{Eq:Unfrustrated}
\Gamma=\sum_{j,k}\left|\left[\bS^{-1}\right]_{jk}\right|, \quad \ket{\Psi_{\bu}}=\frac{1}{\sqrt{\Gamma}}\sum_je^{i\theta_j}\cperp{j} ,
\end{equation}
with angles \(\theta_j\) that satisfy every bond. This covers \(d=2\), where there is no loop, and every real \(\bS\) with nonpositive off-diagonal entries. In the latter case, \(\bS=\mathbf{1}-\mathbf{N}\), where \(\mathbf{N}\) is entrywise nonnegative and, by the Perron--Frobenius theorem, has spectral radius \(1-\lmin<1\). Hence \(\bS^{-1}=\sum_{n\geq0}\mathbf{N}^n\) is entrywise nonnegative~\cite{HornJohnson2013}, and \(\theta_j=0\). For three reference states with equal real overlap \(s\), the case shown in figure~\ref{Fig:Trine}, this covers \(s\leq0\), where \(\Gamma=3/(1+2s)\) and the maximally superposed state is proportional to \(\sum_j\ket{c_j}\).

For \(s\in(0,1)\), in contrast, \([\bS^{-1}]_{jk}=-s/[(1-s)(1+2s)]<0\) for \(j\neq k\). The cyclic product of the three dual overlaps is negative, so the \(XY\) triangle is antiferromagnetic and frustrated. Its ground states are the \(120^{\circ}\) configurations \(\bu=(1,\omega,\omega^2)^{\T}\) and its complex conjugate, with \(\omega=e^{2\pi i/3}\), and the maximally superposed states are the chiral states
\begin{equation}\label{Eq:Chiral}
\ket{\Psi_{\bu}}=\frac{\ket{c_1}+\omega^{\pm1}\ket{c_2}+\omega^{\mp1}\ket{c_3}}{\sqrt{3(1-s)}}, \quad \Gamma=\frac{3}{1-s} .
\end{equation}
For \(s=1/2\), these are exactly the candidates for golden states that Theurer \textit{et al}~\cite{Theurer2017QSup} obtained from the \(\ell_1\) measure of superposition. No superposition with real coefficients reaches this value. Indeed, for a real \(\bpsi\), the argument of equation~\eqref{Eq:CS} with \(\bu\in\{\pm1\}^3\) gives
\begin{equation}\label{Eq:RealTrine}
\max_{\bpsi\in\mathbb{R}^3}\left[1+\MR(\psi)\right]=\max_{\bu\in\{\pm1\}^3}\bu^{\T}\bS^{-1}\bu=\frac{3+5s}{(1-s)(1+2s)}<\frac{3}{1-s},
\end{equation}
as shown in figure~\ref{Fig:Trine}. Frustration thus forces complex coefficients. Since this \(\bS\) is circulant, \(\Gamma=d/\lmin\) nevertheless holds: frustration and the L\"{o}wdin criterion are independent properties.

\begin{figure}[t]
\centering
\includegraphics[width=0.62\textwidth]{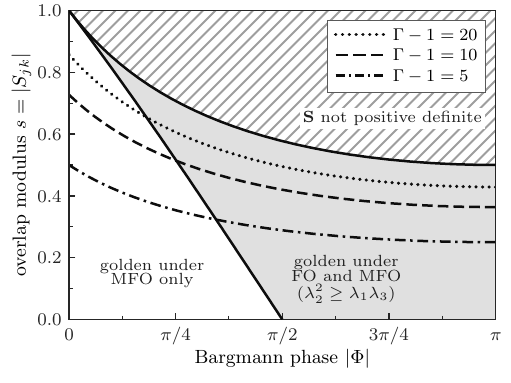}
\caption{Golden states for three reference states with equal overlap modulus \(s\) and Bargmann phase \(\Phi\). Above the upper solid curve (hatched), \(\bS\) is not positive definite. Below it, golden states under MFO exist everywhere, and golden states under FO exist exactly in the shaded region, \(s\geq2\cos[(\pi+|\Phi|)/3]\) (theorem~\ref{Thm:GoldenFO}). Dotted, dashed, and dash-dotted lines: contours of \(\Gamma-1\) from equation~\eqref{Eq:Bargmann}. The edges \(\Phi=0\) and \(|\Phi|=\pi\) correspond to \(s>0\) and, with modulus \(|s|\), to \(s<0\) in figure~\ref{Fig:Trine}.}
\label{Fig:Bargmann}
\end{figure}

More generally, any three states with \(|S_{jk}|=s>0\) and Bargmann phase \(\Phi\coloneqq\arg(S_{12}S_{23}S_{31})\in[-\pi,\pi]\) can be rephased to \(S_{12}=S_{23}=S_{31}=se^{i\Phi/3}\). This matrix is circulant, so its eigenvectors are Fourier vectors of uniform moduli, with the eigenvalues \(1+2s\cos[(\Phi+2\pi k)/3]\), \(k=0,1,2\). Hence \(\Gamma=3/\lmin\), which gives
\begin{equation}\label{Eq:Bargmann}
\Gamma=\frac{3}{1-2s\cos\left[(\pi-|\Phi|)/3\right]} .
\end{equation}
At fixed \(s\), the maximal superposition grows monotonically with \(|\Phi|\), from \(3/(1-s)\) at \(\Phi=0\) to \(3/(1-2s)\) at \(|\Phi|=\pi\) (figure~\ref{Fig:Bargmann}). The geometric phase of the reference states is thus a control parameter of maximal superposition, invisible in the moduli of the overlaps.

In general, the maximization in equation~\eqref{Eq:Gamma} is a nonconvex problem. For \(d\leq3\), it is equivalent to the semidefinite program \(\max\{\Tr(\bS^{-1}\bY):\bY\succeq0,\ Y_{jj}=1\}\) over the complex correlation matrices of lemma~\ref{Lem:Measures}, because for \(d\leq3\) every extreme point of this convex set has rank one; for real correlation matrices, this already fails at \(d=3\). We recall the argument of Li and Tam~\cite{LiTam1994}. Write an extreme point of rank \(r\) as \(\bY=\mathbf{V}\mathbf{V}^{\dagger}\) with \(\mathbf{V}\in\mathbb{C}^{d\times r}\). If \(r^2>d\), some nonzero Hermitian \(r\times r\) matrix \(\mathbf{H}\) satisfies the \(d\) real linear conditions \([\mathbf{V}\mathbf{H}\mathbf{V}^{\dagger}]_{jj}=0\). Then \(\mathbf{V}\mathbf{H}\mathbf{V}^{\dagger}\neq0\), because \(\mathbf{V}\) has rank \(r\), and \(\bY\pm\epsilon\mathbf{V}\mathbf{H}\mathbf{V}^{\dagger}=\mathbf{V}(\mathbf{1}\pm\epsilon\mathbf{H})\mathbf{V}^{\dagger}\) remain correlation matrices for small \(\epsilon\), a contradiction. Hence \(r^2\leq d\), and \(r=1\) for \(d\leq3\). For larger \(d\), this program gives an upper bound on \(\Gamma\), which exceeds \(\Gamma\) by at most the factor \(4/\pi\) because \(\bS^{-1}\) is positive definite~\cite{ZhangHuang2006}. Lower bounds follow from the generalized power method~\cite{Boumal2016}, \(u_j\to[\bS^{-1}\bu]_j/|[\bS^{-1}\bu]_j|\), defined whenever all \([\bS^{-1}\bu]_j\neq0\), which never decreases \(\bu^{\dagger}\bS^{-1}\bu\), by the convexity of this quadratic form. Its fixed points obey \(\bS^{-1}\bu=\mathbf{R}\bu\) with some positive diagonal \(\mathbf{R}\), a weaker condition than equation~\eqref{Eq:Stationary}, so the method can stop below \(\Gamma\).

\section{Gram majorization}
\label{Sec:Gram}

The normalization \(\braket{\psi}{\psi}=\sum_{j,k}\psi_j^{*}S_{jk}\psi_k=1\) of a pure state mixes the populations \(|\psi_j|^2\) of its components \(\psi_j\ket{c_j}\) with their interference. Under FO, the relevant object turns out to be the Gram matrix of these components,
\begin{equation}\label{Eq:GramDef}
\left[\bG_{\psi}\right]_{jk}\coloneqq\psi_j^{*}S_{jk}\psi_k=\left(\psi_j\ket{c_j}\right)^{\dagger}\left(\psi_k\ket{c_k}\right) ,
\end{equation}
that is, \(\bG_{\psi}=\bD_{\psi}^{\dagger}\bS\bD_{\psi}\succeq0\) with \(\bD_{\psi}\coloneqq\mathrm{diag}(\psi_1,\dots,\psi_d)\). Its diagonal holds the populations, its off-diagonal entries are the interference terms, its entries sum to \(\braket{\psi}{\psi}=1\), and its row sums are the CC weights,
\begin{equation}\label{Eq:RowSums}
\bG_{\psi}\mathbf{u}_0=\tilde{\bpsi}, \quad \mathbf{u}_0\coloneqq(1,\dots,1)^{\T} .
\end{equation}
For \(\bS=\mathbf{1}\), \(\bG_{\psi}=\mathrm{diag}(|\psi_1|^2,\dots,|\psi_d|^2)\). For a permutation \(\pi\) of \(\{1,\dots,d\}\), we denote by \(\bP_{\pi}\) the permutation matrix with entries \([\bP_{\pi}]_{ij}\coloneqq\delta_{i,\pi(j)}\), so that \([\bP_{\pi}^{\T}\mathbf{A}\bP_{\pi}]_{jk}=A_{\pi(j)\pi(k)}\) and \([\bP_{\pi}^{\T}\mathbf{y}]_j=y_{\pi(j)}\) for every \(d\times d\) matrix \(\mathbf{A}\) and every vector \(\mathbf{y}\in\mathbb{C}^d\).

\begin{theorem}[Gram majorization]\label{Thm:Gram}
Let \(\ket{\psi}\) be a pure state, and let \(\ket{\varphi}\) be a pure state whose coefficients \(\varphi_j\) are all nonzero. Then \(\ket{\psi}\) can be converted deterministically into \(\ket{\varphi}\) by FO if and only if there is a probability distribution \(\{p_{\pi}\}\) on the permutations of \(\{1,\dots,d\}\) such that
\begin{equation}\label{Eq:GramMaj}
\bG_{\psi}\succeq\sum_{\pi}p_{\pi}\bP_{\pi}^{\T}\bG_{\varphi}\bP_{\pi} .
\end{equation}
\end{theorem}

\begin{proof}
(\(\Rightarrow\)) Let an FO channel with Kraus operators \(\hat{K}_n\) of the form~\eqref{Eq:FOKraus} map \(\ket{\psi}\) to \(\ket{\varphi}\), that is, \(\sum_n\hat{K}_n\ket{\psi}\!\bra{\psi}\hat{K}_n^{\dagger}=\ket{\varphi}\!\bra{\varphi}\). Since a pure state is not a nontrivial mixture, \(\hat{K}_n\ket{\psi}=\gamma_n\ket{\varphi}\) with complex \(\gamma_n\) and \(\sum_n|\gamma_n|^2=1\). If \(\gamma_n=0\), then \(\bK_n\bpsi=0\). If \(\gamma_n\neq0\), every entry of \(\bK_n\bpsi=\gamma_n\bphi\) is nonzero, so \(f_n\) is onto and hence a permutation \(\pi_n\), and \(\kappa_{n,j}\psi_j=\gamma_n\varphi_{\pi_n(j)}\). In particular, all \(\psi_j\) are nonzero, because some \(\gamma_n\) is nonzero. For every \(n\) with \(\gamma_n\neq0\), this gives
\begin{equation}\label{Eq:KSK}
\left[\bK_n^{\dagger}\bS\bK_n\right]_{jk}=\frac{|\gamma_n|^2\left[\bG_{\varphi}\right]_{\pi_n(j)\pi_n(k)}}{\psi_j^{*}\psi_k},
\end{equation}
that is, \(\bD_{\psi}^{\dagger}\bK_n^{\dagger}\bS\bK_n\bD_{\psi}=|\gamma_n|^2\bP_{\pi_n}^{\T}\bG_{\varphi}\bP_{\pi_n}\). Multiplying equation~\eqref{Eq:Completeness} by \(\bD_{\psi}^{\dagger}\) from the left and by \(\bD_{\psi}\) from the right, and grouping the terms with \(\gamma_n\neq0\) by \(\pi_n\), we obtain
\begin{align}
\bG_{\psi}&=\sum_{\pi}p_{\pi}\bP_{\pi}^{\T}\bG_{\varphi}\bP_{\pi}+\bX,\label{Eq:GramDecomp}\\
\bX&\coloneqq\sum_{n:\,\gamma_n=0}\bD_{\psi}^{\dagger}\bK_n^{\dagger}\bS\bK_n\bD_{\psi},\label{Eq:Xdef}
\end{align}
where \(p_{\pi}\coloneqq\sum_{n:\,\pi_n=\pi,\,\gamma_n\neq0}|\gamma_n|^2\) is a probability distribution, because \(\sum_n|\gamma_n|^2=1\). Every term in equation~\eqref{Eq:Xdef} has the form \(\mathbf{A}^{\dagger}\bS\mathbf{A}\succeq0\). Hence \(\bX\succeq0\), and equation~\eqref{Eq:GramMaj} follows.

(\(\Leftarrow\)) The diagonal of equation~\eqref{Eq:GramMaj} gives \(|\psi_j|^2\geq\sum_{\pi}p_{\pi}|\varphi_{\pi(j)}|^2>0\), so all \(\psi_j\) are nonzero. Let \(\bX\succeq0\) be the difference of the two sides of equation~\eqref{Eq:GramMaj}. Because the entries of \(\bG_{\psi}\) and of every \(\bP_{\pi}^{\T}\bG_{\varphi}\bP_{\pi}\) sum to one, \(\mathbf{u}_0^{\T}\bX\mathbf{u}_0=0\). Since \(\bX\succeq0\), this implies \(\bX\mathbf{u}_0=0\), and the spectral decomposition gives \(\bX=\sum_m\mathbf{x}_m\mathbf{x}_m^{\dagger}\) with vectors \(\mathbf{x}_m\) that obey \(\mathbf{x}_m^{\dagger}\mathbf{u}_0=0\). As in~\cite{Torun2021StateTrans} for real coefficients, and in analogy with the weighted permutations used for incoherent operations~\cite{Du2015,ChitambarGour2016PRA}, we define Kraus operators \(\hat{K}_{\pi}\) and \(\hat{L}_m\) of the form~\eqref{Eq:FOKraus} through their matrices on coefficient vectors,
\begin{equation}\label{Eq:KrausConstruct}
\left[\bK_{\pi}\right]_{ij}\coloneqq\sqrt{p_{\pi}}\,\frac{\varphi_{\pi(j)}}{\psi_j}\,\delta_{i,\pi(j)}, \quad \left[\mathbf{L}_m\right]_{ij}\coloneqq\delta_{i1}\frac{\left[\mathbf{x}_m\right]_j^{*}}{\psi_j} ,
\end{equation}
which correspond to \(f=\pi\) and to \(f\equiv1\), respectively. They satisfy \(\bK_{\pi}\bpsi=\sqrt{p_{\pi}}\bphi\) and, because \(\mathbf{x}_m^{\dagger}\mathbf{u}_0=0\), \(\mathbf{L}_m\bpsi=0\). As in equation~\eqref{Eq:KSK}, \(\bD_{\psi}^{\dagger}\bK_{\pi}^{\dagger}\bS\bK_{\pi}\bD_{\psi}=p_{\pi}\bP_{\pi}^{\T}\bG_{\varphi}\bP_{\pi}\). The only nonzero row of \(\mathbf{L}_m\bD_{\psi}\) is its first row \(\mathbf{x}_m^{\dagger}\), so \(S_{11}=1\) gives \(\bD_{\psi}^{\dagger}\mathbf{L}_m^{\dagger}\bS\mathbf{L}_m\bD_{\psi}=\mathbf{x}_m\mathbf{x}_m^{\dagger}\). Hence
\begin{equation}\label{Eq:CompletenessCheck}
\bD_{\psi}^{\dagger}\left(\sum\nolimits_{\pi}\bK_{\pi}^{\dagger}\bS\bK_{\pi}+\sum\nolimits_m\mathbf{L}_m^{\dagger}\bS\mathbf{L}_m\right)\bD_{\psi}=\bG_{\psi},
\end{equation}
which, since \(\bD_{\psi}\) is invertible, is equation~\eqref{Eq:Completeness}. These Kraus operators therefore form an FO channel, which maps \(\ket{\psi}\!\bra{\psi}\) to \(\sum_{\pi}p_{\pi}\ket{\varphi}\!\bra{\varphi}=\ket{\varphi}\!\bra{\varphi}\).
\end{proof}

Three remarks place theorem~\ref{Thm:Gram} in context. First, for \(\bS=\mathbf{1}\), both sides of equation~\eqref{Eq:GramMaj} are diagonal with unit trace, so the inequality becomes the equality \((|\psi_j|^2)_j=\sum_{\pi}p_{\pi}\bP_{\pi}^{\T}(|\varphi_j|^2)_j\). By the theorems of Hardy, Littlewood, and P\'{o}lya and of Birkhoff~\cite{MarshallOlkin2011}, such \(p_{\pi}\) exist if and only if \((|\psi_j|^2)_j\) is majorized by \((|\varphi_j|^2)_j\). This is the criterion for incoherent operations~\cite{Du2015,Du2017Erratum,Zhu2017} and for strictly incoherent operations~\cite{Winter2016,ChitambarGour2016PRA,ChitambarGour2017Erratum}, the analog of Nielsen's theorem for entanglement~\cite{Nielsen1999}. Gram majorization is thus the RTS form of majorization: the populations are replaced by the Gram matrix, and the equality between population vectors by a semidefinite inequality, whose slack \(\bX\) is realized by the Kraus operators \(\hat{L}_m\), which annihilate \(\ket{\psi}\) and have no counterpart in the RTC. Kraus operators that annihilate \(\ket{\psi}\) were used in~\cite{Torun2021StateTrans} to satisfy the completeness relation; here, they are constructed explicitly from \(\bX\).

Second, the matrix \(\bX\) in equation~\eqref{Eq:GramDecomp} obeys \(\bX\mathbf{u}_0=0\), because \(\bD_{\psi}\mathbf{u}_0=\bpsi\) and \(\bK_n\bpsi=0\) whenever \(\gamma_n=0\). Applying equation~\eqref{Eq:GramDecomp} to \(\mathbf{u}_0\) and using equation~\eqref{Eq:RowSums} and \(\bP_{\pi}\mathbf{u}_0=\mathbf{u}_0\) therefore gives the row-sum condition
\begin{equation}\label{Eq:FirstMoment}
\tilde{\bpsi}=\sum_{\pi}p_{\pi}\bP_{\pi}^{\T}\tilde{\bphi}:
\end{equation}
the CC weights of the initial state are a mixture of permutations of those of the target. For real coefficients and real overlaps, conditions of this kind were derived in~\cite{Torun2021StateTrans}: a majorization of the CC weights, supplemented by a condition on the populations. The necessary condition \(\tilde{\psi}_j=1/d\) for golden states found in~\cite{Senyasa2022Golden} is of the same type. For \(d=2\), equation~\eqref{Eq:FirstMoment} and the diagonal of equation~\eqref{Eq:GramMaj} already imply equation~\eqref{Eq:GramMaj} for the same \(\{p_{\pi}\}\): the difference \(\bX\) of the two sides is then a Hermitian \(2\times2\) matrix with \(\bX\mathbf{u}_0=0\), hence a real multiple of \((1,-1)^{\T}(1,-1)\), and its nonnegative diagonal gives \(\bX\succeq0\). In general, theorem~\ref{Thm:Gram} completes these conditions by the full matrix inequality, which constrains every interference term, including its phase.

Relative phases are constrained as well. The phase rotation \(\sum_je^{i\eta_j}\ket{c_j}\!\bra{c_j^{\perp}}\), which maps each \(\ket{c_j}\) to \(e^{i\eta_j}\ket{c_j}\), has the form~\eqref{Eq:FOKraus} with \(f\) the identity, so it is an operation in FO if and only if it is unitary, that is, if it satisfies equation~\eqref{Eq:Completeness}: \(e^{i(\eta_k-\eta_j)}S_{jk}=S_{jk}\) for all \(j,k\). Unlike in the RTC, where every phase rotation is free, only global phases remain when all overlaps are nonzero. Rephasing the reference states, as in section~\ref{Sec:Frustration}, is a different matter: it changes neither the free states nor the free operations, only the phase convention of the vectors \(\ket{c_j}\).

Third, the hypothesis on \(\ket{\varphi}\) cannot be dropped. Every state can be converted into the free state \(\ket{c_1}\) by FO, for instance with the Kraus operators \(\ket{c_1}\!\bra{l_j}\), \(j=1,\dots,d\), which have the form~\eqref{Eq:FOKraus} with \(f\equiv1\). For the target \(\bphi=\mathbf{e}_1\coloneqq(1,0,\dots,0)^{\T}\), however, equation~\eqref{Eq:GramMaj} implies \(\bX\mathbf{u}_0=0\), as in the proof of sufficiency, and hence \(\tilde{\bpsi}=\sum_{\pi}p_{\pi}\bP_{\pi}^{\T}\mathbf{e}_1\), which fails unless all CC weights of \(\ket{\psi}\) are real and nonnegative. The proof of sufficiency, in contrast, holds for every target once all \(\psi_j\) are nonzero. For given states, equation~\eqref{Eq:GramMaj} defines a semidefinite feasibility problem in the \(d!\) variables \(p_{\pi}\), which is readily solvable for small \(d\).

\section{Golden states under FO}
\label{Sec:GoldenFO}

We now apply Gram majorization to golden states under FO. Theorem~\ref{Thm:GoldenFO} gives a necessary condition, part (a), which extends that of~\cite{Senyasa2022Golden} to all of FO, and a sufficient condition, part (b). Parts (c) and (d) decide the existence question completely for two families of overlap matrices, and corollary~\ref{Cor:Three} below settles the case of three reference states.

\begin{theorem}[Golden states under FO]\label{Thm:GoldenFO}
(a) If a golden state under FO exists, then \(\Gamma(\bS)=d/\lmin\), and every golden state under FO is a ground state of \(\hat{\chi}\) that is maximally coherent in \(\mathcal{L}\). In terms of \(\mathcal{C}\), such a state reads
\begin{equation}\label{Eq:GoldenFO}
\ket{\Psi}=\frac{1}{\sqrt{d\lmin}}\sum_{j=1}^{d}u_j\ket{c_j},
\end{equation}
with \(\bu\in\mathbb{T}^d\) and \(\bS\bu=\lmin\bu\).\\
(b) If \(\bS\) has exactly two distinct eigenvalues and \(\lmin\) is nondegenerate, then \(\lmin\) has an eigenvector \(\bu\in\mathbb{T}^d\), and the state~\eqref{Eq:GoldenFO} is golden under FO.\\
(c) If \(d\geq3\) and \(S_{jk}=s\in(0,1)\) for all \(j\neq k\), no golden state under FO exists.\\
(d) If \(d=3\) and \(|S_{jk}|=s\in(0,1)\) for all \(j\neq k\), with Bargmann phase \(\Phi=\arg(S_{12}S_{23}S_{31})\in[-\pi,\pi]\), a golden state under FO exists if and only if
\begin{equation}\label{Eq:TrineCriterion}
s\geq2\cos\frac{\pi+|\Phi|}{3},
\end{equation}
or, equivalently, if and only if the eigenvalues \(\lambda_1\leq\lambda_2\leq\lambda_3\) of \(\bS\) obey \(\lambda_2^2\geq\lambda_1\lambda_3\).
\end{theorem}

\begin{proof}
(a) A golden state under FO is golden under MFO (section~\ref{Sec:Free}), so, by theorem~\ref{Thm:Maximal}(c), it is some \(\ket{\Psi_{\bu}}\) with optimal \(\bu\), whose coefficients are all nonzero. We first show that some target with nonzero coefficients has uniform CC weights. As in the diagonal scaling of positive definite matrices~\cite{MarshallOlkin1968}, the function \(\sum_j\ln|\varphi_j|\), which tends to \(-\infty\) as any coefficient vanishes, attains its maximum on the compact set \(\bphi^{\dagger}\bS\bphi=1\) at a point with nonzero coefficients. The Lagrange condition there reads \(\varphi_j^{*}[\bS\bphi]_j=\mathrm{const}\), and the constant equals \(1/d\) by equation~\eqref{Eq:CC}. Applying equation~\eqref{Eq:FirstMoment} to the conversion of \(\ket{\Psi_{\bu}}\) into this \(\ket{\varphi}\) gives \(\tilde{\Psi}_j=1/d\). Since \(\tilde{\Psi}_j=r_j/\Gamma\) by equation~\eqref{Eq:Stationary}, all \(r_j\) equal \(\Gamma/d\). Hence \(\bS\bu=\lambda\bu\) with \(\lambda=d/\Gamma\), and the coefficients of \(\ket{\Psi_{\bu}}\) are \(u_j/\sqrt{d\lambda}\).

It remains to show that \(\lambda=\lmin\). Consider the targets \(\ket{\varphi_\epsilon}\propto\ket{c_1}+\epsilon\sum_{k\geq2}\ket{c_k}\), whose coefficients are nonzero for \(\epsilon\neq0\). Theorem~\ref{Thm:Gram} and equation~\eqref{Eq:FirstMoment} provide probability distributions \(\{p^{\epsilon}_{\pi}\}\) with
\begin{equation}\label{Eq:EpsTargets}
\bG_{\Psi}\succeq\sum_{\pi}p^{\epsilon}_{\pi}\bP_{\pi}^{\T}\bG_{\varphi_\epsilon}\bP_{\pi}, \quad \frac{\mathbf{u}_0}{d}=\sum_{\pi}p^{\epsilon}_{\pi}\bP_{\pi}^{\T}\tilde{\bphi}_\epsilon ,
\end{equation}
where \(\bG_{\Psi}\) is the Gram matrix of \(\ket{\Psi_{\bu}}\). As \(\epsilon\to0\), \(\bG_{\varphi_\epsilon}\to\mathbf{e}_1\mathbf{e}_1^{\T}\) and \(\tilde{\bphi}_\epsilon\to\mathbf{e}_1\). Along a subsequence, \(p^{\epsilon}_{\pi}\to p_{\pi}\) for every \(\pi\), and in the limit \(\sum_{\pi}p_{\pi}\bP_{\pi}^{\T}\mathbf{e}_1=\mathbf{u}_0/d\). Since every \(\bP_{\pi}^{\T}\mathbf{e}_1\) is a standard basis vector, this gives \(\bG_{\Psi}\succeq\sum_{\pi}p_{\pi}\bP_{\pi}^{\T}\mathbf{e}_1\mathbf{e}_1^{\T}\bP_{\pi}=\mathbf{1}/d\). With the unitary \(\bD_{\bu}\coloneqq\mathrm{diag}(u_1,\dots,u_d)\), \(\bG_{\Psi}=\bD_{\bu}^{\dagger}\bS\bD_{\bu}/(d\lambda)\), so this means \(\bS\succeq\lambda\mathbf{1}\), that is, \(\lmin\geq\lambda\), and hence \(\lambda=\lmin\). Thus \(\Gamma=d/\lmin\), the coefficients of \(\ket{\Psi_{\bu}}\) are those of equation~\eqref{Eq:GoldenFO}, and, by proposition~\ref{Prop:Lowdin}(a), \(\ket{\Psi_{\bu}}\) is a ground state of \(\hat{\chi}\) that is maximally coherent in \(\mathcal{L}\).

(b) Since \(\Tr\bS=d\), the other eigenvalue, of multiplicity \(d-1\), is \(\mu=(d-\lmin)/(d-1)\), and \(\bS=\mu\mathbf{1}-(\mu-\lmin)\mathbf{v}\mathbf{v}^{\dagger}\) with a normalized eigenvector \(\mathbf{v}\) for \(\lmin\). The unit diagonal of \(\bS\) gives \(|v_j|^2=(\mu-1)/(\mu-\lmin)=1/d\), so \(\bu\coloneqq\sqrt{d}\,\mathbf{v}\in\mathbb{T}^d\), and \(\bS=\mu\mathbf{1}-(\mu-\lmin)\bu\bu^{\dagger}/d\). With the all-ones matrix \(\mathbf{J}\coloneqq\mathbf{u}_0\mathbf{u}_0^{\T}\), the state~\eqref{Eq:GoldenFO} has the Gram matrix
\begin{equation}\label{Eq:GPsiB}
\bG_{\Psi}=\frac{1}{d\lmin}\left[\mu\mathbf{1}-\frac{\mu-\lmin}{d}\mathbf{J}\right].
\end{equation}
Let \(\ket{\varphi}\) have nonzero coefficients, let \(\|\bphi\|^2\coloneqq\sum_j|\varphi_j|^2\), and choose \(p_{\pi}=1/d!\), as in~\cite{Senyasa2022Golden}. Since \(\bG_{\varphi}\) has trace \(\|\bphi\|^2\) and entries summing to one,
\begin{equation}\label{Eq:Symmetrized}
\frac{1}{d!}\sum_{\pi}\bP_{\pi}^{\T}\bG_{\varphi}\bP_{\pi}=\frac{\|\bphi\|^2}{d}\mathbf{1}+\frac{1-\|\bphi\|^2}{d(d-1)}(\mathbf{J}-\mathbf{1}) .
\end{equation}
Both sides of equation~\eqref{Eq:GramMaj} are then combinations of \(\mathbf{1}\) and \(\mathbf{J}\), and both have \(\mathbf{u}_0\) as an eigenvector with eigenvalue \(1/d\). On the orthogonal complement of \(\mathbf{u}_0\), equation~\eqref{Eq:GramMaj} therefore reads \(\mu/(d\lmin)\geq(d\|\bphi\|^2-1)/[d(d-1)]\). This holds because \(\|\bphi\|^2\leq\bphi^{\dagger}\bS\bphi/\lmin=1/\lmin\) and \((d/\lmin-1)/(d-1)=\mu/\lmin\). By theorem~\ref{Thm:Gram}, \(\ket{\Psi}\) can thus be converted into every pure state with nonzero coefficients. The set of FO channels is compact: their Choi matrices are exactly the sums, over the \(d^d\) functions \(f\), of positive semidefinite matrices supported on fixed subspaces, one for each \(f\) (the eigenvectors of each summand give Kraus operators of the form~\eqref{Eq:FOKraus}), subject to the linear condition of trace preservation, which also fixes the total trace. The set of states reachable from \(\ket{\Psi}\) is therefore closed, and it contains all pure states, because those with nonzero coefficients are dense among them. It is also convex, because the set of FO channels is convex, and hence it contains all states.

(c) Here \(\bS=(1-s)\mathbf{1}+s\mathbf{J}\), so \(\lmin=1-s\), and the eigenvectors for \(\lmin\) are the nonzero vectors orthogonal to \(\mathbf{u}_0\). By part (a), a golden state would therefore be a state~\eqref{Eq:GoldenFO} with \(\bu\in\mathbb{T}^d\) and \(\sum_ju_j=0\). Fix such a \(\bu\), and take a target \(\ket{\varphi}\) with nonzero coefficients, \(\sum_j\varphi_j=0\), and nonuniform moduli, for example \(\bphi\propto(1,\dots,1,1-d)^{\T}\). Then \(\bS\bphi=(1-s)\bphi\), so \(\bphi^{\dagger}\bS\bphi=(1-s)\|\bphi\|^2\) and \(\|\bphi\|^2=1/(1-s)\). Suppose that equation~\eqref{Eq:GramMaj} holds with some \(p_{\pi}\), and let \(\overline{\bG}_{\varphi}\coloneqq\sum_{\pi}p_{\pi}\bP_{\pi}^{\T}\bG_{\varphi}\bP_{\pi}\), so that \(\Tr\overline{\bG}_{\varphi}=\|\bphi\|^2\). The candidate has \(\bG_{\Psi}=\bD_{\bu}^{\dagger}\bS\bD_{\bu}/[d(1-s)]=[(1-s)\mathbf{1}+s\,\bu^{*}\bu^{\T}]/[d(1-s)]\), which acts as \(\mathbf{1}/d\) on the orthogonal complement of \(\bu^{*}\). Hence the trace of \(\overline{\bG}_{\varphi}\) over this \((d-1)\)-dimensional subspace obeys
\begin{equation}\label{Eq:TraceBound}
\|\bphi\|^2-\frac{\bu^{\T}\overline{\bG}_{\varphi}\bu^{*}}{d}\leq\frac{d-1}{d} .
\end{equation}
For each \(\pi\), \(\bu^{\T}\bP_{\pi}^{\T}\bG_{\varphi}\bP_{\pi}\bu^{*}=\|\sum_j\varphi_j'\ket{c_j}\|^2\) with the rephased coefficients \(\varphi_j'\coloneqq[\bP_{\pi}\bu^{*}]_j\varphi_j\), which obey \(|\varphi_j'|=|\varphi_j|\). Moreover, \(\|\sum_j\varphi_j'\ket{c_j}\|^2=(1-s)\sum_j|\varphi_j'|^2+s|\sum_j\varphi_j'|^2\leq(1-s)\|\bphi\|^2+s\|\bphi\|_1^2\), where \(\|\bphi\|_1\coloneqq\sum_j|\varphi_j|\). Averaging over \(p_{\pi}\) and using equation~\eqref{Eq:TraceBound}, we obtain \(d\|\bphi\|^2-(d-1)\leq\bu^{\T}\overline{\bG}_{\varphi}\bu^{*}\leq(1-s)\|\bphi\|^2+s\|\bphi\|_1^2\), and with \(\|\bphi\|^2=1/(1-s)\), this yields \(\|\bphi\|_1^2\geq d/(1-s)=d\|\bphi\|^2\). This contradicts the Cauchy--Schwarz inequality, which is strict for nonuniform moduli. Hence the candidate cannot be converted into \(\ket{\varphi}\). Since \(\bu\) was arbitrary, no golden state under FO exists.

(d) For \(\Phi=0\), the overlaps can be rephased to \(S_{jk}=s\), so part (c) applies; equation~\eqref{Eq:TrineCriterion} fails as well, because \(2\cos(\pi/3)=1>s\). Let \(0<|\Phi|\leq\pi\). As in section~\ref{Sec:Frustration}, a rephasing gives \(S_{12}=S_{23}=S_{31}=se^{i\Phi/3}\). Then \(\bS\) is circulant, with the Fourier eigenvectors \(\bff_k\coloneqq(1,\omega^k,\omega^{2k})^{\T}/\sqrt{3}\), \(\omega=e^{2\pi i/3}\), \(k\in\mathbb{Z}_3\), and the eigenvalues \(1+2s\cos[(\Phi+2\pi k)/3]\). The smallest eigenvalue \(\lambda_1\) is nondegenerate, because two of the angles \((\Phi+2\pi k)/3\) share the smallest cosine only if they are \(2\pi/3\) and \(4\pi/3\) modulo \(2\pi\), which requires \(\Phi=0\). Hence part (a) leaves, up to a global phase, a single candidate, \(\sum_ju_j\ket{c_j}/\sqrt{3\lambda_1}\), where \(\bu/\sqrt{3}\) is the Fourier vector for \(\lambda_1\). The further rephasing \(\ket{c_j}\to u_j\ket{c_j}\) keeps \(\bS\) circulant and turns the candidate into \(\ket{\Psi}=\sum_j\ket{c_j}/\sqrt{3\lambda_1}\). From now on, \(\bS\) denotes the overlap matrix after both rephasings, and \(\nu_k\) its eigenvalue for \(\bff_k\). Thus \(\nu_0=\lambda_1\), \(\{\nu_1,\nu_2\}=\{\lambda_2,\lambda_3\}\), and \(\bG_{\Psi}=\bS/(3\lambda_1)\).

Fix a target \(\ket{\varphi}\) with nonzero coefficients, and suppose that equation~\eqref{Eq:GramMaj} holds for some \(p_{\pi}\). Reading the labels modulo \(3\), every permutation of \(\{1,2,3\}\) has the form \(j\mapsto\pm j+n\) with \(n\in\mathbb{Z}_3\). The even ones are the cyclic shifts \(\tau_n\colon j\mapsto j+n\), which leave the circulant matrices \(\bS\) and \(\bG_{\Psi}\) invariant, \(\bP_{\tau_n}^{\T}\bS\bP_{\tau_n}=\bS\). Conjugating equation~\eqref{Eq:GramMaj} with the matrices \(\bP_{\tau_n}\) shows that it also holds for each of the distributions \(\pi\mapsto p_{\pi\circ\tau_n}\), and hence, by linearity in \(p_{\pi}\), for their average over \(n\). This average is constant on the even and on the odd permutations, so we may assume that \(p_{\pi}=q/3\) for even \(\pi\) and \(p_{\pi}=(1-q)/3\) for odd \(\pi\), with \(q\in[0,1]\). The right side of equation~\eqref{Eq:GramMaj} is then invariant under conjugation with the shifts, so it is circulant and diagonal in the Fourier basis, like \(\bG_{\Psi}\). Since every permutation has the form \(j\mapsto\pm j+n\), \(\bP_{\pi}\bff_k\) is proportional to \(\bff_k\) for even \(\pi\) and to \(\bff_{-k}\) for odd \(\pi\). With \(g_k\coloneqq\bff_k^{\dagger}\bG_{\varphi}\bff_k\), equation~\eqref{Eq:GramMaj} is thus equivalent to
\begin{equation}\label{Eq:TrineModes}
\frac{\nu_k}{3\lambda_1}\geq q\,g_k+(1-q)\,g_{-k}, \quad k=1,2;
\end{equation}
for \(k=0\), both sides equal \(1/3\), because \(\bff_0=\mathbf{u}_0/\sqrt{3}\). Write \(\bphi=\sum_m\beta_m\bff_m\). The normalization \(\bphi^{\dagger}\bS\bphi=\sum_m\nu_m|\beta_m|^2=1\) makes \(\zeta_m\coloneqq\nu_m|\beta_m|^2\) a probability vector. Moreover, \(g_k=\|\sum_j[\bff_k]_j\varphi_j\ket{c_j}\|^2\), and, since \(\sqrt{3}\,[\bff_k]_j[\bff_m]_j=[\bff_{m+k}]_j\), this superposition has the coefficient vector \(\sum_m\beta_m\bff_{m+k}/\sqrt{3}\), so that
\begin{equation}\label{Eq:TrineG}
3g_k=\sum_{m\in\mathbb{Z}_3}\zeta_m\frac{\nu_{m+k}}{\nu_m} .
\end{equation}

A value \(q\in[0,1]\) that satisfies equation~\eqref{Eq:TrineModes} exists if and only if \(g_1+g_2\leq(\nu_1+\nu_2)/(3\lambda_1)\) and \(\min(g_1,g_2)\leq\min(\nu_1,\nu_2)/(3\lambda_1)\), because the right sides of equation~\eqref{Eq:TrineModes} for \(k=1,2\) add up to \(g_1+g_2\), and each of them runs through the interval between \(g_1\) and \(g_2\) as \(q\) varies. The first condition always holds: by equation~\eqref{Eq:TrineG} and \(\Tr\bS=3\), \(3(g_1+g_2)=\sum_m\zeta_m(3-\nu_m)/\nu_m\leq(3-\lambda_1)/\lambda_1=(\nu_1+\nu_2)/\lambda_1\). For the second, let \(\nu_1\leq\nu_2\) (otherwise exchange the roles of \(\bff_1\) and \(\bff_2\)). By equation~\eqref{Eq:TrineG}, \(3g_1\) is a mixture of \(x_0\coloneqq\nu_1/\lambda_1\), \(x_1\coloneqq\nu_2/\nu_1\), and \(x_2\coloneqq\lambda_1/\nu_2\leq1\leq x_0\). If \(x_1\leq x_0\), then \(g_1\leq\nu_1/(3\lambda_1)\) for every target. If \(x_1>x_0\), then \(\nu_1<\nu_2\), and the targets with \((\zeta_0,\zeta_1,\zeta_2)=(1-\epsilon,\epsilon,0)\) have \(3g_1>x_0\) and, for small \(\epsilon>0\), \(3g_2>x_0\), because \(3g_2\to\nu_2/\lambda_1>x_0\); their coefficients \(\beta_0[\bff_0]_j+\beta_1[\bff_1]_j\) are nonzero, because \(|\beta_1|<|\beta_0|\) for small \(\epsilon\). With the closure argument of part (b), a golden state therefore exists if and only if \(x_1\leq x_0\), that is, \(\nu_1^2\geq\lambda_1\nu_2\), which reads \(\lambda_2^2\geq\lambda_1\lambda_3\) because \(\nu_1=\lambda_2\) and \(\nu_2=\lambda_3\). Finally, the eigenvalues are \(\lambda_1=1-2sA\), \(\lambda_2=1-2sB\), and \(\lambda_3=1+2s(A+B)\), with \(A\coloneqq\cos[(\pi-|\Phi|)/3]\) and \(B\coloneqq\cos[(\pi+|\Phi|)/3]\). Because \(A^2+AB+B^2=3/4\),
\begin{equation}\label{Eq:TrineIdentity}
\lambda_2^2-\lambda_1\lambda_3=3s\,(s-2B),
\end{equation}
which proves equation~\eqref{Eq:TrineCriterion}.
\end{proof}

Theorem~\ref{Thm:GoldenFO}(a) derives, from theorem~\ref{Thm:Maximal}(c) and Gram majorization, the necessary conditions of~\cite{Senyasa2022Golden}, an eigenvector of \(\bS\) for \(\lmin\) with uniform CC weights, for all of FO; in~\cite{Senyasa2022Golden}, they were obtained for a restricted family of Kraus operators. Part (a) also proves, for every golden state under FO, what was found in~\cite{Torun2023Golden} for the golden states known then: they are maximally coherent in the L\"{o}wdin basis. This is a property of golden states under FO, not of maximal superposition as such. Under MFO, golden states exist for every \(\bS\) by theorem~\ref{Thm:Maximal}(c), but, by proposition~\ref{Prop:Lowdin}(b), they are maximally coherent in the L\"{o}wdin basis only if they are eigenstates of \(\hat{\chi}\), which fails in general (figure~\ref{Fig:LowdinWeight}).

Part (b) covers all pairs of nonorthogonal reference states, for which golden states were known~\cite{Theurer2017QSup,Torun2021StateTrans}. For \(d\geq3\), it is the rephasing-invariant form of the configurations for which golden states were found in~\cite{Senyasa2022Golden}: after the rephasing \(\ket{c_j}\to u_j\ket{c_j}\), all overlaps equal \(-(1-\lmin)/(d-1)\), and the golden state becomes the equal-weight superposition of the rephased reference states.

For \(d=3\) and \(s=1/2\), Theurer \textit{et al}~\cite{Theurer2017QSup} took the candidates for golden states to have coefficients of equal modulus, by the symmetry of the reference states, singled out the chiral states~\eqref{Eq:Chiral} by maximizing the \(\ell_1\) measure, and excluded them by a semidefinite bound on the probability of reaching a single target; the candidates with real coefficients were excluded in~\cite{Senyasa2022Golden}. Part (c) extends this result to all \(s\in(0,1)\) and all \(d\geq3\), and part (a) shows that the candidates must indeed have coefficients of equal modulus. For \(d=3\), the chiral states satisfy all conditions of part (a), so these conditions, conjectured in~\cite{Senyasa2022Golden} to be sufficient as well, are not sufficient, as the example of~\cite{Theurer2017QSup} already indicates.

The obstruction in part (c) comes from the off-diagonal interference terms alone. With \(p_{\pi}=1/d!\), the candidates and the targets used in the proof satisfy both the row-sum condition~\eqref{Eq:FirstMoment} and the diagonal of equation~\eqref{Eq:GramMaj}, because the CC weights \((1-s)|\varphi_j|^2\) of these targets sum to one and \(\|\bphi\|^2/d=1/[d(1-s)]=|\Psi_j|^2\). Hence, for \(d\geq3\), equation~\eqref{Eq:FirstMoment} and the diagonal of equation~\eqref{Eq:GramMaj} do not imply equation~\eqref{Eq:GramMaj}, in contrast to \(d=2\) (section~\ref{Sec:Gram}).

Since rephasing does not affect the existence of golden states, part (c) also covers every \(\bS\) with \(d\geq3\) and exactly two distinct eigenvalues of which the larger is nondegenerate: the unit diagonal of \(\bS\) then gives the corresponding eigenvector entries of equal modulus, and a rephasing makes all overlaps equal to \(1-\lmin>0\). Together with part (b), theorem~\ref{Thm:GoldenFO} thus decides every \(\bS\) with exactly two distinct eigenvalues of which at least one is nondegenerate: a golden state under FO exists if and only if \(\lmin\) is nondegenerate. When both eigenvalues are degenerate, which requires \(d\geq4\), the question remains open.

Part (d) settles the existence question for three reference states with equal overlap moduli; see figure~\ref{Fig:Bargmann}. Equation~\eqref{Eq:TrineCriterion} can be read as
\begin{equation}\label{Eq:Threshold}
|\Phi|\geq3\arccos\frac{s}{2}-\pi .
\end{equation}
Golden states under FO therefore exist for all positive definite \(\bS\) of this family when \(|\Phi|\geq\pi/2\), never when \(\Phi=0\), and in between only for sufficiently strong overlaps; as \(s\to0\), the threshold phase tends to \(\pi/2\). For \(0<|\Phi|<\pi\), the three eigenvalues of \(\bS\) are distinct, so this region lies outside part (b), which contains only \(|\Phi|=\pi\). The two real cases are thus the extremes: \(\Phi=0\) never admits golden states under FO, and \(|\Phi|=\pi\) always does. For \(0<|\Phi|<\pi\), the Bargmann invariant is not real, so the reference states cannot be given real amplitudes in any common basis; they form an imaginary set in the basis-independent sense of~\cite{Fernandes2024} (for the resource theory of imaginarity, see~\cite{HickeyGour2018,Wu2021}). In this regime, the existence of golden states under FO depends, like \(\Gamma\) (section~\ref{Sec:Frustration}), on the geometric phase \(\Phi\), which the overlap moduli do not determine. Throughout this family, \(\bS\) is circulant after rephasing, so \(\Gamma=3/\lmin\), and the golden states under MFO, which exist for all \(s\) and \(\Phi\), are maximally coherent in the L\"{o}wdin basis.

For three reference states, theorem~\ref{Thm:GoldenFO} decides the existence question completely, because part (a) already forces the overlap moduli to be equal, as the proof below shows. This confirms, for \(d=3\), the observation in~\cite{Senyasa2022Golden} that unequal overlap moduli in general exclude golden states.

\begin{corollary}[Three reference states]\label{Cor:Three}
For \(d=3\), a golden state under FO exists if and only if the reference states are orthonormal or \(|S_{jk}|=s\in(0,1)\) for all \(j\neq k\) with equation~\eqref{Eq:TrineCriterion} satisfied. Equivalently, it exists if and only if \(\lmin\) has an eigenvector in \(\mathbb{T}^3\) and the eigenvalues \(\lambda_1\leq\lambda_2\leq\lambda_3\) of \(\bS\) obey \(\lambda_2^2\geq\lambda_1\lambda_3\).
\end{corollary}

\begin{proof}
By theorem~\ref{Thm:GoldenFO}(a), a golden state requires \(\bS\bu=\lmin\bu\) for some \(\bu\in\mathbb{T}^3\). The rephasing \(\ket{c_j}\to u_j\ket{c_j}\) changes neither the overlap moduli nor the eigenvalues and makes \(\mathbf{v}_1\coloneqq\mathbf{u}_0/\sqrt{3}\) an eigenvector for \(\lambda_1=\lmin\). Complete it to an orthonormal eigenbasis \(\mathbf{v}_1,\mathbf{v}_2,\mathbf{v}_3\) with eigenvalues \(\lambda_1\leq\lambda_2\leq\lambda_3\). The unit diagonal of \(\bS\) and the completeness of the eigenbasis give, for each \(j\),
\begin{equation}\label{Eq:ThreeRows}
\sum_{k=2,3}\lambda_k\left|[\mathbf{v}_k]_j\right|^2=1-\frac{\lambda_1}{3}, \quad \sum_{k=2,3}\left|[\mathbf{v}_k]_j\right|^2=\frac{2}{3} .
\end{equation}
If \(\lambda_2\neq\lambda_3\), these equations fix \(|[\mathbf{v}_2]_j|^2\) independently of \(j\), so it equals \(1/3\). Up to a phase, \(\sqrt{3}\,\mathbf{v}_2\) is then a phase vector with first entry \(1\), and orthogonality to \(\mathbf{v}_1\) makes its three entries sum to zero. Three unit complex numbers with zero sum form an equilateral triangle, so \(\sqrt{3}\,\mathbf{v}_2\) is \((1,\omega,\omega^2)^{\T}\) or \((1,\omega^2,\omega)^{\T}\). Hence \(\mathbf{v}_2\) and \(\mathbf{v}_3\) are Fourier vectors, \(\bS\) is circulant, and its off-diagonal entries share a common modulus. If \(\lambda_2=\lambda_3\), then \(\bS=\lambda_2\mathbf{1}-(\lambda_2-\lambda_1)\mathbf{u}_0\mathbf{u}_0^{\T}/3\), whose off-diagonal entries all equal \((\lambda_1-\lambda_2)/3\). In both cases, the overlaps have a common modulus \(s\). For \(s=0\), the reference states are orthonormal, and golden states exist; for \(s\in(0,1)\), theorem~\ref{Thm:GoldenFO}(d) applies. The spectral form follows because a common modulus makes \(\bS\) circulant after rephasing, with Fourier eigenvectors, and equation~\eqref{Eq:TrineCriterion} is equivalent to \(\lambda_2^2\geq\lambda_1\lambda_3\).
\end{proof}

In particular, three reference states with real nonpositive overlaps admit a golden state under FO if and only if all overlaps are equal, and the family of equation~\eqref{Eq:Isosceles} admits one if and only if \(a=b\), in which case its maximally superposed state is golden under FO and maximally coherent in the L\"{o}wdin basis (figure~\ref{Fig:LowdinWeight}).

\section{Witnessing superposition in the L\"{o}wdin basis}
\label{Sec:Witness}

We now turn the equal-overlap states into witnesses of superposition. Free states are in general themselves coherent in the L\"{o}wdin basis, so a useful witness must separate the coherence that free states carry from the excess due to superposition. For \(\bu\in\mathbb{T}^d\), let \(\ket{\xi_{\bu}}\coloneqq\sum_ju_j\cperp{j}\), so that \(\ket{\Psi_{\bu}}=\ket{\xi_{\bu}}/\sqrt{\Gamma}\) for optimal \(\bu\), and define the observable \(\hat{W}_{\bu}\coloneqq\ket{\xi_{\bu}}\!\bra{\xi_{\bu}}\). We also define the phase-weighted L\"{o}wdin coherence
\begin{equation}\label{Eq:LowdinCoherence}
\mathcal{K}_{\bu}(\hat{\sigma})\coloneqq\sum_{j\neq k}u_j^{*}\left[\bsig_{\mathrm{L}}\right]_{jk}u_k .
\end{equation}
Since the L\"{o}wdin weights sum to one, \(\bu^{\dagger}\bsig_{\mathrm{L}}\bu=1+\mathcal{K}_{\bu}(\hat{\sigma})\). This number lies in \([0,d]\), because \(\bsig_{\mathrm{L}}\succeq0\) has unit trace and \(\|\bu\|^2=d\), so \(\mathcal{K}_{\bu}(\hat{\sigma})\) is real and lies in \([-1,d-1]\).

\begin{proposition}[L\"{o}wdin witnesses]\label{Prop:Witness}
(a) For every state, \(\Tr(\hat{W}_{\bu}\hat{\sigma})=\bu^{\dagger}\bsig\bu\). In particular, \(\Tr(\hat{W}_{\bu}\hat{\delta})=1\) for every free state \(\hat{\delta}\) and every \(\bu\in\mathbb{T}^d\), and \(\hat{\sigma}\) is free if and only if \(\Tr(\hat{W}_{\bu}\hat{\sigma})=1\) for all \(\bu\in\mathbb{T}^d\).\\
(b) For every state, \(1+\MR(\hat{\sigma})\geq\max_{\bu\in\mathbb{T}^d}\Tr(\hat{W}_{\bu}\hat{\sigma})\), with equality for pure states and, if \(d\leq3\), for all states.\\
(c) If \(\Gamma=d/\lmin\) and \(\bu\) is optimal, then \(\hat{W}_{\bu}=(d/\lmin)\ket{\Psi_{\bu}}\!\bra{\Psi_{\bu}}\) with \(\braket{l_j}{\Psi_{\bu}}=u_j/\sqrt{d}\), and
\begin{equation}\label{Eq:WitnessLowdin}
\Tr(\hat{W}_{\bu}\hat{\sigma})=\frac{1+\mathcal{K}_{\bu}(\hat{\sigma})}{\lmin} .
\end{equation}
Every free state has \(\mathcal{K}_{\bu}=\lmin-1\), and every state obeys
\begin{equation}\label{Eq:WitnessBound}
\lmin\,\MR(\hat{\sigma})\geq\mathcal{K}_{\bu}(\hat{\sigma})-(\lmin-1),
\end{equation}
with equality for \(\ket{\Psi_{\bu}}\).
\end{proposition}

\begin{proof}
(a) By equation~\eqref{Eq:State}, \(\Tr(\hat{W}_{\bu}\hat{\sigma})=\sum_{j,k}u_j^{*}u_k\bra{c_j^{\perp}}\hat{\sigma}\cperp{k}=\bu^{\dagger}\bsig\bu\). For a free state, \(\bsig\) is diagonal with \(\Tr\bsig=\Tr(\bsig\bS)=1\), so this equals one. Conversely, \(\bu^{\dagger}\bsig\bu=\sum_j\sigma_{jj}+\sum_{j\neq k}u_j^{*}u_k\sigma_{jk}\) is a trigonometric polynomial in the phases \(\theta_j\) of \(u_j=e^{i\theta_j}\), with the coefficient \(\sigma_{jk}\) at the frequency \(e^{i(\theta_k-\theta_j)}\). It is constant only if \(\sigma_{jk}=0\) for all \(j\neq k\). Then \(\bsig\succeq0\) is diagonal with \(\Tr\bsig=\Tr(\bsig\bS)=1\), so \(\hat{\sigma}\) is free.

(b) The inequality is equation~\eqref{Eq:RobSDP} restricted to the correlation matrices \(\bY=\bu\bu^{\dagger}\) with \(\bu\in\mathbb{T}^d\). For a pure state \(\ket{\psi}\), the phases \(u_j=\psi_j/|\psi_j|\), with \(u_j=1\) if \(\psi_j=0\), give \(\Tr(\hat{W}_{\bu}\ket{\psi}\!\bra{\psi})=|\bu^{\dagger}\bpsi|^2=(\sum_j|\psi_j|)^2\), which equals \(1+\MR(\psi)\) by equation~\eqref{Eq:PureMeasures}. For \(d\leq3\), every extreme point of the set of correlation matrices has rank one (section~\ref{Sec:Frustration}), that is, has the form \(\bu\bu^{\dagger}\) with \(\bu\in\mathbb{T}^d\), and a linear function attains its maximum over a compact convex set at an extreme point.

(c) Since \(\bu^{\dagger}\bS^{-1}\bu=\Gamma=d/\lmin=\|\bu\|^2/\lmin\), the Rayleigh bound of equation~\eqref{Eq:Bounds} is attained, so \(\bS\bu=\lmin\bu\). By equation~\eqref{Eq:Bases}, \(\ket{\xi_{\bu}}\) then has the coefficients \(\bS^{-1/2}\bu=\bu/\sqrt{\lmin}\) in \(\mathcal{L}\). Hence \(\braket{l_j}{\Psi_{\bu}}=u_j/\sqrt{\lmin\Gamma}=u_j/\sqrt{d}\), \(\hat{W}_{\bu}=\Gamma\ket{\Psi_{\bu}}\!\bra{\Psi_{\bu}}=(d/\lmin)\ket{\Psi_{\bu}}\!\bra{\Psi_{\bu}}\), and \(\Tr(\hat{W}_{\bu}\hat{\sigma})=\bra{\xi_{\bu}}\hat{\sigma}\ket{\xi_{\bu}}=\bu^{\dagger}\bsig_{\mathrm{L}}\bu/\lmin\), which is equation~\eqref{Eq:WitnessLowdin}, because \(\bu^{\dagger}\bsig_{\mathrm{L}}\bu=1+\mathcal{K}_{\bu}(\hat{\sigma})\). The value of \(\mathcal{K}_{\bu}\) for free states follows from part (a), and equation~\eqref{Eq:WitnessBound} from part (b). For \(\ket{\Psi_{\bu}}\), \(\mathcal{K}_{\bu}=d-1\) by equation~\eqref{Eq:WitnessLowdin}, and \(\MR=\Gamma-1=d/\lmin-1\) by theorem~\ref{Thm:Maximal}(b), so both sides of equation~\eqref{Eq:WitnessBound} equal \(d-\lmin\).
\end{proof}

\begin{figure}[t]
\centering
\includegraphics[width=0.62\textwidth]{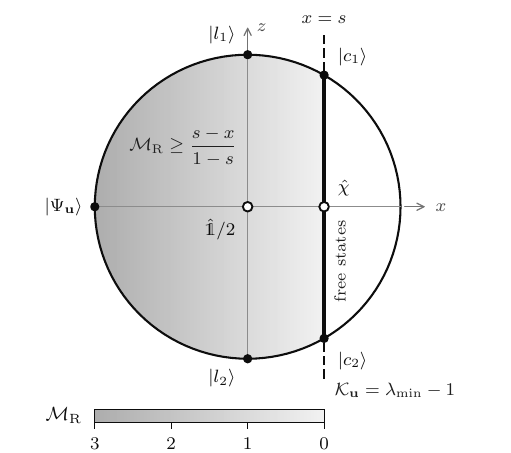}
\caption{L\"{o}wdin witness for \(d=2\) and \(S_{12}=s=1/2\), shown in the disk of real states (\(y=0\)) of the Bloch ball, with the L\"{o}wdin basis \(\ket{l_1}\), \(\ket{l_2}\) at the poles. All free states lie on the thick chord, which lies in the plane \(x=s\) (dashed), where the L\"{o}wdin coherence \(\mathcal{K}_{\bu}=-x\) for the optimal \(\bu=(1,-1)^{\T}\) equals \(\lmin-1\). Every state off this plane carries superposition; in this slice, all states in the plane are free. In the shaded region, equation~\eqref{Eq:WitnessBound} gives \(\MR\geq(s-x)/(1-s)\); this bound is attained throughout the shaded region and is largest at the golden state \(\ket{\Psi_{\bu}}\). The maximally mixed state (center) carries superposition; the uniform mixture \(\hat{\chi}\) does not.}
\label{Fig:Witness}
\end{figure}

Proposition~\ref{Prop:Witness} answers the third question raised in section~\ref{Sec:Intro}. Consider the case \(\Gamma=d/\lmin\) of part (c). The L\"{o}wdin weights of a state drop out of equation~\eqref{Eq:WitnessLowdin}, since they always sum to one, and the information sits in the L\"{o}wdin coherences. For example, the maximally superposed state \(\ket{\Psi_{\bu}}\), the uniform mixture \(\hat{\chi}\), and the maximally mixed state \(\hat{\Id}/d\) all have \(w_j=1/d\), whereas \(\mathcal{K}_{\bu}\) takes the values \(d-1\), \(\lmin-1\), and \(0\), respectively. For an optimal \(\bu\), the L\"{o}wdin coherence that the overlaps imprint on the free states takes the single value \(\lmin-1\leq0\) for every free state. Subtracting it leaves the excess L\"{o}wdin coherence \(\mathcal{K}_{\bu}-(\lmin-1)\), which certifies superposition whenever it is nonzero and, by equation~\eqref{Eq:WitnessBound}, bounds the robustness from below when it is positive. For \(d=2\), figure~\ref{Fig:Witness} shows the geometry: the free states fill a chord of the Bloch ball, which lies in the plane on which \(\mathcal{K}_{\bu}\) takes its free-state value \(\lmin-1\).

A simple consequence concerns white noise. The maximally mixed state \(\hat{\Id}/d\) on \(\mathcal{H}\) has \(\bsig=\bS^{-1}/d\), uniform L\"{o}wdin weights, and no L\"{o}wdin coherence, so \(\mathcal{K}_{\bu}(\hat{\Id}/d)=0\). Proposition~\ref{Prop:Witness}(b) gives a lower bound on its robustness, and equation~\eqref{Eq:RobSDP}, together with \(\Tr(\bS^{-1}\bY)\leq\Tr\bY/\lmin=d/\lmin\), gives an upper bound:
\begin{equation}\label{Eq:WhiteNoise}
\frac{\Gamma}{d}-1\leq\MR(\hat{\Id}/d)\leq\frac{1}{\lmin}-1,
\end{equation}
with equality on the left for \(d\leq3\), and on both sides when \(\Gamma=d/\lmin\). Moreover, by equation~\eqref{Eq:RobSDP}, \(d[1+\MR(\hat{\Id}/d)]\) is exactly the semidefinite relaxation of \(\Gamma\) discussed in section~\ref{Sec:Frustration}, so that also \(\MR(\hat{\Id}/d)\leq4\Gamma/(\pi d)-1\). By equation~\eqref{Eq:Bounds}, the lower bound is positive for every \(\bS\neq\mathbf{1}\). White noise, which is free in the RTC, is thus a superposition resource (figure~\ref{Fig:Trine}). Its L\"{o}wdin weights coincide with those of the free state \(\hat{\chi}\), and, when \(\Gamma=d/\lmin\), its excess L\"{o}wdin coherence, \(1-\lmin\), reveals the resource.

In the usual terminology~\cite{Napoli2016,Regula2018}, both \(\hat{W}_{\bu}-\hat{\Id}\) and \(\hat{\Id}-\hat{W}_{\bu}\) are superposition witnesses: by part (a), their expectation values vanish on every free state, and together they detect every resource state for a suitable \(\bu\). For pure states, and for all states when \(d\leq3\), \(\hat{\Id}-\hat{W}_{\bu}\) alone suffices, because part (b) then gives \(\max_{\bu}\Tr(\hat{W}_{\bu}\hat{\sigma})=1+\MR(\hat{\sigma})>1\) for every resource state. For \(\bS=\mathbf{1}\), \(\hat{W}_{\bu}=d\ket{\Psi_{\bu}}\!\bra{\Psi_{\bu}}\) with a maximally coherent state \(\ket{\Psi_{\bu}}\), and the inequality in part (b) is an instance of the witness bound on the robustness of coherence~\cite{Napoli2016}.

For \(d\geq4\), the set of correlation matrices has extreme points of rank two~\cite{LiTam1994}, and the inequality in proposition~\ref{Prop:Witness}(b) can be strict for mixed states. An example is the state whose coefficient matrix is proportional to the Gram matrix \(\mathbf{Q}\) of four single-qubit states whose Bloch vectors form a regular tetrahedron. Since \(\mathbf{Q}\) is a correlation matrix with the eigenvalues \(2,2,0,0\), the choice \(\bY=\mathbf{Q}\) is optimal in equation~\eqref{Eq:RobSDP}, with \(\Tr\mathbf{Q}^2=8\), whereas \(\max_{\bu}\bu^{\dagger}\mathbf{Q}\bu=2(2+\sqrt{3})\). The ratio of the two sides of the inequality is therefore \(4/(2+\sqrt{3})\approx1.07\), independently of \(\bS\). For \(\bS=(1-\epsilon)\mathbf{Q}+\epsilon\mathbf{1}\) with \(0<\epsilon<1-\sqrt{3}/2\), the same state even has \(\max_{\bu}\Tr(\hat{W}_{\bu}\hat{\sigma})<1\), although \(\MR(\hat{\sigma})=\epsilon/(2-\epsilon)>0\), so that only \(\hat{W}_{\bu}-\hat{\Id}\) detects it.

Experimentally, the witness for an optimal \(\bu\), \(\hat{W}_{\bu}=\Gamma\ket{\Psi_{\bu}}\!\bra{\Psi_{\bu}}\), requires a single two-outcome measurement, \(\{\ket{\Psi_{\bu}}\!\bra{\Psi_{\bu}},\hat{\Id}-\ket{\Psi_{\bu}}\!\bra{\Psi_{\bu}}\}\), also when \(\Gamma<d/\lmin\). By equation~\eqref{Eq:PsiUprops}, every free state yields the first outcome with the same probability,
\begin{equation}\label{Eq:FreeProbability}
\bra{\Psi_{\bu}}\hat{\delta}\ket{\Psi_{\bu}}=\frac{1}{\Gamma} \quad \text{for all}\ \hat{\delta}\in\Fc,
\end{equation}
and any deviation, in either direction, certifies superposition.

For the phase-shifted coherent states of equation~\eqref{Eq:Cat}, this measurement has a direct physical form. Each residue class of the photon number modulo \(d\) intersects the span of \(\mathcal{C}\) in the one-dimensional space spanned by a single cat state. On this span, the projector onto \(\ket{\Psi_{\bu}}\) therefore coincides with the projector onto the photon numbers \(n\equiv m\pmod d\), with the residue \(m\) of equation~\eqref{Eq:Cat}, and the witness is a measurement of the photon number modulo \(d\). Such a measurement has been proposed for error detection in rotation-symmetric bosonic codes~\cite{Grimsmo2020}; for \(d=2\), it is the photon-number parity, which is measured repeatedly and nondestructively in circuit quantum electrodynamics~\cite{Sun2014}. Every free state yields the residue \(m\) with probability \(\lmin/d=1/\Gamma\), the probability of that residue in a coherent state. The identification holds for states in the span of \(\mathcal{C}\); photon loss, for instance, damps the amplitude \(\alpha\) and thereby leads out of this span, and the residue statistics of such states do not refer to superposition relative to \(\mathcal{C}\).

\section{Conclusion}
\label{Sec:Conclusion}

We have shown what decides the existence of golden states in the RTS and when they are maximally coherent in the L\"{o}wdin basis. The answer depends on the free operations. Under MFO, golden states exist for every overlap matrix, and they are exactly the maximally superposed states. Under FO, every golden state is maximally coherent in the L\"{o}wdin basis, but golden states exist only for special overlap matrices, and for three reference states a geometric phase decides whether they exist. The L\"{o}wdin pattern of the known golden states~\cite{Torun2023Golden} is therefore a property of FO, not of maximal superposition.

The robustness and the geometric measure of superposition are maximized by the same states. They are the superpositions \(\sum_ju_j\cperp{j}\) of the dual basis vectors with optimal phases, which overlap equally with all reference states, and a single number, \(\Gamma(\bS)=\max_{\bu\in\mathbb{T}^d}\bu^{\dagger}\bS^{-1}\bu\), fixes both maximal values and the largest advantage that any state offers in channel discrimination. This number is set by the ground-state energy of a classical \(XY\) model on the dual overlaps. Frustration in this model can force complex coefficients even when all overlaps are real, and \(\Gamma\) depends on the geometric phases of the reference states, not only on the moduli of their overlaps. The maximally superposed states are maximally coherent in the L\"{o}wdin basis exactly when \(\Gamma=d/\lmin\).

Under FO, Gram majorization solves the conversion problem for pure states: a pure state can be converted into a target with nonzero coefficients if and only if the Gram matrices of their components obey a semidefinite ordering up to a mixture of permutations. For orthonormal reference states, it reduces to the majorization criterion of the RTC; for nonorthogonal ones, it constrains every interference term, including its phase. Applied to golden states, it shows that every golden state under FO is a ground state of \(\hat{\chi}\) that is maximally coherent in the L\"{o}wdin basis, so that \(\Gamma=d/\lmin\) is necessary, but not sufficient. If \(\bS\) has exactly two distinct eigenvalues, at least one of them nondegenerate, golden states under FO exist if and only if \(\lmin\) is nondegenerate; in particular, for \(d\geq3\) reference states with equal real overlaps, they exist when the overlaps are negative and never when they are positive. For three reference states that are not orthonormal, golden states under FO exist if and only if the overlaps have a common modulus \(s\) and the Bargmann phase obeys \(s\geq2\cos[(\pi+|\Phi|)/3]\). Up to rephasing, \(\Phi=0\) and \(|\Phi|=\pi\) are the cases of equal positive and equal negative overlaps; between these two real extremes, the geometric phase decides whether the theory has a unit.

Finally, the equal-overlap states provide a complete family of superposition witnesses in the L\"{o}wdin basis, although the free states are themselves coherent in this basis. When \(\Gamma=d/\lmin\), all free states carry the same phase-weighted L\"{o}wdin coherence, \(\lmin-1\); any deviation from this value certifies superposition, and a positive excess bounds the robustness from below. Each witness requires a single two-outcome measurement, which, for phase-shifted coherent states, is a measurement of the photon number modulo \(d\), the photon-number parity for \(d=2\). As a consequence, white noise on the span of the reference states, which is free in the RTC, is a superposition resource, with robustness at least \(\Gamma/d-1\), whenever the reference states are not orthonormal.

These results replace the observed L\"{o}wdin pattern by exact statements: \(\Gamma(\bS)\) quantifies maximal superposition, Gram majorization governs the conversion of pure states under FO, and, for three reference states, the Bargmann phase decides whether golden states exist.

\section*{Declaration on the use of artificial intelligence}

In the course of this work, the author used Claude Opus 5.5 (Anthropic) to assist in refining the mathematical proofs, writing the code for the numerical checks and the figures, and preparing the manuscript. The central idea, connecting LSO with the RTC and the RTS, is the author's own and goes back to his earlier work~\cite{Torun2023Golden}; he also directed the present research. The author checked all statements, derivations, and references and takes full responsibility for the content of this publication.

\section*{References}

\bibliographystyle{iopart-num}
\bibliography{refs}

\end{document}